\documentclass[twocolumn,aps,prb,
10pt,
amssymb,superscriptaddress]{revtex4-2}

\usepackage{hyperref}
\usepackage{amsmath}
\usepackage{amsfonts}
\usepackage{graphicx}
\usepackage{appendix}
\usepackage{dsfont}
\usepackage{amssymb}
\usepackage{bm}
\usepackage{xcolor}
\definecolor{plum}{rgb}{0.56, 0.27, 0.52}
\definecolor{tealmodern}{rgb}{0.0, 0.5, 0.5}
\usepackage{soul}
\usepackage{natbib}
\setstcolor{red}
\usepackage{nccmath}
\usepackage{array}
\usepackage{footmisc}
\usepackage{comment}

\usepackage[english]{babel}
\newcommand{\beq}{\begin{equation*}}
\newcommand{\eneq}{\end{equation*}}
\mathchardef\mhyphen="2D

\usepackage{xcolor}
\usepackage{hyperref}
\usepackage[mathlines]{lineno}
\usepackage[normalem]{ulem}
\usepackage{xr}

\definecolor{modernorange}{HTML}{FF6B35}
\definecolor{darkblue}{HTML}{0066CC}
\definecolor{tealmodern}{HTML}{006D77}
\definecolor{slate}{HTML}{37474F}  
\definecolor{plum}{HTML}{6A4C93}

\begin{document}

\title{Exact fluctuation relations in voltage- and temperature-biased\\ Laughlin-edge constrictions} 

\author{Gu Zhang}
\affiliation{National Laboratory of Solid State Microstructures, School of Physics, Jiangsu Physical Science Research Center and Collaborative Innovation Center of Advanced Microstructures, Nanjing University, Nanjing 210093, China}

\author{Gabriele Campagnano}
\affiliation{CNR-SPIN, c/o Complesso di Monte S. Angelo, via Cinthia - 80126 - Napoli, Italy}

\author{Domenico Giuliano}
\affiliation{Institut für Theoretische Physik, Heinrich-Heine-Universität, 40225 Düsseldorf, Germany}
\affiliation{INFN, Gruppo collegato di Cosenza, Arcavacata di Rende, I-87036 Cosenza, Italy}
\affiliation{Dipartimento di Fisica, Università della Calabria, Arcavacata di Rende, I-87036 Cosenza, Italy}

\author{Igor Gornyi}
\affiliation{Institute for Quantum Materials and Technologies and Institut für Theorie der Kondensierten Materie, Karlsruhe Institute of Technology, 76131 Karlsruhe, Germany}

\author{Inès Safi}
\affiliation{Laboratoire de Physique des Solides (UMR 5802), CNRS-Université Paris-Sud and Paris-Saclay, Bâtiment 510, 91405 Orsay, France}

\begin{abstract}

We present a comprehensive analysis of non-equilibrium fluctuation-dissipation relations connecting experimentally accessible chiral-current auto- and cross-correlations to the tunneling-current noise and conductance in Laughlin edge states coupled through a quantum point contact (QPC). We examine their validity for two chiral Laughlin edges held at different temperatures and voltages and show that the relations remain exact for arbitrary tunneling strength, voltage bias, and edge-state temperatures. We further generalize them to spatially extended QPCs and to tunneling amplitudes with an explicit voltage dependence, and discuss the conditions and limitations associated with these generalizations. Our results establish that the local tunneling-current noise generated at the QPC can be reliably reconstructed from experimentally accessible auto- and cross-correlations measured downstream, providing a robust route to characterize non-equilibrium transport in chiral edge states.

\end{abstract}

\maketitle

\section{Introduction}

Quantum transport, encompassing charge and heat currents and their associated
noises, has emerged as a powerful tool for probing and manipulating exotic
condensed-matter phenomena \cite{IhnBook,BelzigPhysicsToday10}. It is
particularly effective for revealing the properties of topological edge states
in fractional quantum Hall (FQH) systems, where quantized transport plateaus
provide signatures of topological order
\cite{KlitzingRevModPhys86,StormerTsuiGossardRMP99,HasanKaneRMP10,
MaciejkoHughesZhangARCMP11}. Out-of-equilibrium shot noise of weak tunneling
currents through quantum point contacts provides direct access to quasiparticle
properties: fractional charges via Poissonian noise in Laughlin states
\cite{Kane:1994b,ines_resonance,HeiblumNature97,SaminadayarGlattiJinEtiennePRL97}
and signatures of fractional statistics through current correlations
\cite{Safi:2001,Deviatov2012,Campagnano:2012,Campagnano:2013,fractional_statistics_theory_2016,
NakamuraNatPhys19,fractional_statistics_gwendal_science_2020,
NakamuraNatPhys20,LeeNature23,HeiblumNP2023,ghosh2024,ines_alex_2025}. Extending these
determinations to higher-order Abelian
\cite{
Kane:1994,heiblum_varying_frac_PRB_2010,Bid:2010,NosigliaParkRosenowGefenPRB18,SpanslattParkGefenMirlinPRL19,NakamuraNatPhys19,ParkMirlinRosenowGefenPRB19,SpanslattParkGefenMirlinPRB20,ParkRosenowGefenPRR21,SpanslattGefenGornyiPolyakovPRB21,SchillerOregSnizhkoPRB22,pierre_anyons_PRX_2023,
fractional_statistics_gwendal_PRX_2023,HanLeeSimPRL24,ParkGefenPRB24,MannaDasGefenGoldsteinLTP24,SrivastavDasModPhysLett25}
and non-Abelian FQH edges
\cite{Bid:2010,LafontScience19,ParkSpanslattGefenMirlinPRL20,SchillerOregSnizhkoPRB22,DuttaScience22,HanLeeSimPRL24,ChouDasSarmaPRB24,AlkalayX2026}
remains challenging. These systems have nevertheless attracted increasing
attention, in part because of their possible relevance for topological quantum
technologies.

Despite theoretical advances in microscopic modeling of low-energy edge-state
theories
\cite{wen_review_FQHE_1992,jain_FQHE_compiste_fermions_PRL_1989,
kane_fisher_jain}
with local QPCs, including Bethe-Ansatz solutions
\cite{Fendley:1995} and refermionization approaches that extend beyond the
weak-tunneling regime while preserving locality, a mismatch with most
experiments persists. At the same time, the simplest Tomonaga--Luttinger
scaling for local weak tunneling has recently found experimental confirmation
\cite{manfra_TLL_graphene_2025,benjamin_zimmermann_TLL_7_3_9_PhD_2016};
see also Ref.~\cite{VeillonNature24} and the related discussion in
Ref.~\cite{kyrylo_FQHE_2024}. The experimental situation therefore motivates
two closely related goals: deriving robust transport relations that depend as
little as possible on microscopic details of strongly correlated systems, and
extending their validity beyond the idealized assumptions of a local QPC,
energy-independent tunneling, or weak tunneling.

A natural starting point for such model-independent relations is the fluctuation-dissipation theorem (FDT).
The FDT is a cornerstone of equilibrium
statistical physics, establishing a fundamental relation between fluctuations
and response in thermal equilibrium. Through detailed balance, its connection
to the Kubo formula in the linear-response regime is well established.
At finite drives, however, the equilibrium FDT no longer applies in its usual
form, motivating the search for non-equilibrium fluctuation-dissipation
relations (FDRs). Extensive studies have established such relations for
zero-frequency noise in weakly nonlinear and near-equilibrium regimes
\cite{FDT_tobiska_nazarov_PRB_2005,FDT_PRB_2008,butt_FDR_PRL_2008,
FDT_Schon_PRB_2010,FDT_eq_nonlinear_kobayashi_PRB_11,
FDT_revue_RevModPhys_2009,FDT_gaspard_multi_NJP_2013,WangFeldmanPRL13,
FDT_broken_TR_symmetry_feldman_IJMPB_2014,
FDT_multi_spin_currents_feldman_PRB_2015}. Such results, however, do not by
themselves cover arbitrary strongly driven stationary states or the full range
of transport situations involving non-equilibrium reservoirs, temperature
gradients, weakly coupled circuit elements, or engineered thermal environments
\cite{various_T_thermal_machine_PRE_2017,various_T_TLL_lebowitz_PRB_2017,
ines_pierre_thermal_2021,hekking_various_T_PRB_2013,
various_T_thermal_noise_entanglement_PRA_2017}.

Two complementary strategies address these limitations without requiring
microscopic details. First, the unifying non-equilibrium perturbative approach
\cite{ines_eugene,ines_cond_mat,ines_PRB_2019,ines_photo_noise_PRB_2022}
has established common transport relations across a broad range of systems,
including integer and fractional quantum Hall devices, normal and Josephson
junctions, thermal transport setups, and anyonic colliders. The same framework
can accommodate temperature gradients and has been used, for example, to
formulate non-invasive thermoelectric probes
\cite{SafiPRB20}. It has also enabled robust determinations of fractional
charge beyond the Laughlin series through photo-assisted
\cite{ines_eugene,ines_cond_mat,ines_PRB_2019,ines_photo_noise_PRB_2022}
and finite-frequency noise
\cite{BenaSafiPBR07,ines_degiovanni_2016,SafiPRB20,ines_gwendal},
as well as proposals for accessing fractional statistics
\cite{SafiPRB20,ines_statistics_2025};
see also Ref.~\cite{fractional_statistics_zhang_Gefen_2025_anyon_collider}
for the \textit{non-thermal} case.

Second, a non-perturbative framework, directly relevant to the present work,
establishes exact relations between finite-frequency noise asymmetries and
generalized non-equilibrium admittances for arbitrary stationary states,
time-dependent drives, and interacting multi-terminal systems
\cite{SafiBenaCrepieuxPRB08,ines_proceedings_2009,ines_philippe}.
While exact FDRs were first obtained for linear systems
\cite{LesovikJETP1997} and later for nonlinear systems under more restrictive
Hamiltonian assumptions \cite{GavishPRB00,Gavish04}, subsequent developments
have led to Kubo-type relations beyond thermal equilibrium
\cite{meir,nonlinear_kubo_EPL_2010,FDT_nonlinear_kubo_Svirskii_JETP_1981}
and to a unified description of non-equilibrium phenomena including
noise asymmetries, excess-noise effects, and negative excess noise
\cite{LesovikJETP1997,deblock_06,cottet_08,dolcini_07,
Lesovik_loosen_1993_negative,
fractional_statistics_Sim_PRL_negative_shot_noise_PRL_2019}.
These methods have potential applications to a wide range of mesoscopic systems
\cite{zamoum_12,BenaSafiPBR07,admittance_FF_kondo_andergassen_PRB_13,
admittance_FF_dot_wolfle_PRB_13,simon_nonlinear_zarand_2011,
simon_non_linear_response_theory_2014,
FDT_nonlinear_quantum_dots_ding_PRB_2013,
FDT_nonlinear_Ness_dash_PRL_2012,FDT_nonlinear_volker_PRB_2022,
wang_ADN_non_linear_response_2015,nazarov_back_action_FDR_cite_PRB2019}
and Josephson-type circuits
\cite{FDT_nonlinear_JJ_joyez_PRX_2020,
FDT_nonlinear_JJ_dupuis_PRB_2023,
FDT_nonlinear_JJ_yuto_PRL_2022,
FDT_nonlinear_JJ_luca_PRB_2024}.

The exact FDR for DC-voltage noise asymmetry is, in particular, relevant to both chiral current correlations and tunneling noise in Hall edges containing a QPC.
Here, an important practical issue arises. Theoretical studies of quantum transport naturally focus on the tunneling current and its fluctuations, namely the tunneling-current noise generated locally at the QPC.
Experiments, however, generally measure auto- and cross-correlations of currents collected at drains located downstream and far from the QPC. Extracting the local tunneling-current noise from these observables therefore requires knowledge of the relation between the measured drain currents and the tunneling processes
at the QPC~\cite{ines_gwendal,fractional_statistics_gwendal_science_2020}.
This issue becomes especially important out of equilibrium, where the vicinity
of the QPC may involve dynamically injected quasiparticles, interactions, and
equilibration processes, while experimentally accessible current detectors are
placed well outside this region. Relations between the local tunneling-current
noise and downstream auto- and cross-correlations thus provide the necessary
bridge between quantities naturally calculated in theory and quantities
directly measured in experiment.

In the context of FQH edge transport, such relations have been discussed
extensively. Perturbative relations apply when quasiparticle tunneling is weak or when the system is near the opposite strong-coupling limit
\cite{fractional_statistics_theory_2016}. Beyond perturbative treatments, exact
relations between tunneling-current noise and chiral current noise have been
established for zero-frequency auto-correlations
\cite{Kane:1994b,Fendley:1996,WangFeldmanPRL13,FeldmanMotyPRB17}.
More generally, generalized non-equilibrium response identities have been used
to derive fluctuation relations in bosonized out-of-equilibrium
backscattering models.
In these settings, a one-dimensional conductor described by a bosonized theory is perturbed by a local or spatially extended backscattering/tunneling operator, and current correlations are related to
tunneling-current noise and differential response. This formulation has been
applied both to nonchiral Luttinger-liquid geometries with impurity
backscattering
\cite{DolciniPRB05,SafiBenaCrepieuxPRB08}
and to chiral Laughlin-edge constrictions
\cite{BenaSafiPBR07,zamoum_12}. The formalism can also accommodate spatially
extended scattering regions \cite{SafiBenaCrepieuxPRB08}.

More recently, a strict relation between tunneling-current noise and both
auto- and cross-correlations was established for a system consisting of two
channels of a $\nu=1/2$ chiral Luttinger liquid carrying quasiparticles with
fractional charge $e/2$ \cite{OneHalfPRB24}. Previous explicit applications of
these relations to Hall-edge tunneling have, however, usually assumed an
energy-independent tunneling amplitude at the QPC. Such an assumption may be too restrictive, as FQH tunneling experiments have shown a range of behavior, with some results departing from the simplest chiral-Luttinger-liquid scaling predictions and others remaining consistent with them \cite{kyrylo_FQHE_2024,VeillonNature24}.
Moreover, in the previous explicit applications
\cite{Kane:1994b,Fendley:1996,DolciniPRB05,BenaSafiPBR07,
SafiBenaCrepieuxPRB08,zamoum_12,WangFeldmanPRL13,FeldmanMotyPRB17,
OneHalfPRB24},
the two channels involved are usually taken to have the same temperature.
Allowing different edge temperatures is essential for situations involving
temperature gradients and is central to the rapidly developing field of
delta-$T$ noise
\cite{LumbrosoNature18,SafiPRB20,RechJonckheereGremaudMartinPRL20,
ErikssonPRL21,DeltaTPRB22,MatteoSciPhys25}.

In this work, we address these limitations by investigating a system in which two
Laughlin edges with different voltages and temperatures communicate through an
extended QPC. The scattering region may have a finite spatial extent, and the
effective transfer amplitude need not be momentum independent. As our main result, we establish, for stationary systems hosting chiral channels (Fig.~\ref{fig:model}), an exact, non-perturbative zero-frequency fluctuation relation [Eq.~\eqref{eq:noise_conductance_relation}] between the tunneling-current noise through the QPC and experimentally accessible current correlations, including both auto- and cross-correlations, which remains valid arbitrarily far from equilibrium.

This non-equilibrium fluctuation relation, which can be viewed as an important
zero-frequency manifestation of more general fluctuation-response identities
\cite{DolciniPRB05,SafiBenaCrepieuxPRB08,BenaSafiPBR07}, remains valid beyond
the perturbative tunneling regime. More importantly for experimentally
realistic settings, we show explicitly that it survives several
generalizations that are absent from most previous Hall-edge applications:
a spatially extended QPC with a nontrivial momentum dependence of the transfer
amplitude
(Sec.~\ref{sec:momentum} and
Appendix~\ref{app:momentum_dependent}),
a temperature difference between the two channels (see Fig.~\ref{fig:heat_flow}), provided that the size of the scattering region is much smaller than the equilibration length but much larger than the relevant thermal diffusion length, and a
tunneling amplitude that depends explicitly on the applied bias
(Sec.~\ref{sec:voltage_dependence}).

These generalizations extend previous results
\cite{Kane:1994b,Fendley:1996,BenaSafiPBR07,zamoum_12,
WangFeldmanPRL13,FeldmanMotyPRB17,OneHalfPRB24}
to experimentally relevant situations involving temperature gradients and
nontrivial spatial, momentum, or bias dependence of the transfer process.
The theoretical framework required to establish the relation under these
conditions is itself an important part of the work and builds on the methods of
Refs.~\cite{DolciniPRB05,BenaSafiPBR07,ines_proceedings_2009,ines_philippe}.
The resulting relation provides a direct route for obtaining the locally generated tunneling-current noise, which is not measured directly, from auto- and cross-correlations measured far from the QPC. Its validity for channels at different temperatures also makes the relation  applicable to delta-$T$ noise experiments.

The rest of the work is organized as follows.
We begin in Sec.~\ref{sec:system} by introducing the system and stating the
general non-equilibrium fluctuation relation,
Eq.~\eqref{eq:noise_conductance_relation}.
In Sec.~\ref{sec:generalized_linear_response}, we develop the theoretical
framework
\cite{DolciniPRB05,BenaSafiPBR07,ines_proceedings_2009,ines_philippe}
used throughout the paper.
Sec.~\ref{sec:noise_conductance_relation} contains the detailed derivation
of Eq.~\eqref{eq:noise_conductance_relation}.
The result is extended in Sec.~\ref{sec:voltage_dependence} to systems where
the tunneling amplitude depends explicitly on the applied voltage biases.
Finally, Sec.~\ref{sec:summary} summarizes the results and discusses the range
of applicability of the non-equilibrium fluctuation relation.

\begin{widetext}
\begin{table*}[ht]
\begin{center}
\begin{tabular}{ |c|c|c| } 
 \hline
 Symbols & Physical meaning & Definition \\ 
 \hline
 $\phi_\alpha$ & Edge-$\alpha$ bosonic operator 
 & Eqs.~\eqref{eq:lead_hamiltonian} and \eqref{eq:bosonization} \\ 
 \hline
 $V_\alpha$ & Voltage bias of edge $\alpha$ 
 & Eq.~\eqref{eq:hbias} \\ 
 \hline
 $V$ & Voltage-bias difference 
 & Above Eq.~\eqref{eq:hbias} \\ 
 \hline
 $G$ & Tunneling charge conductance 
 & Eqs.~\eqref{eq:noise_conductance_relation_brief} and 
   \eqref{eq:noise_conductance_relation} \\ 
 \hline
 $\hat{I}_\mathcal{T}$ & Tunneling current operator 
 & Eq.~\eqref{eq:iud_general} \\ 
 \hline
 $\hat{I}_\alpha$ & Current operator of edge-$\alpha$ 
 & Eq.~\eqref{eq:charge_current} \\ 
 \hline
 $S_\mathcal{T}$ & Tunneling current noise 
 & Eq.~\eqref{eq:tunneling_current_noise} \\ 
 \hline
 $S_{\alpha\alpha'}$ & Correlation between $\hat{I}_\alpha$ and $\hat{I}_{\alpha'}$ 
 & Eq.~\eqref{eq:auto_cross_correlations} \\ 
 \hline
 $\hat{Q}_\alpha$ & Charge operator of edge-$\alpha$ 
 & Eq.~\eqref{eq:charge_operator} \\ 
 \hline
 $\hat{Q}_{\text{QPC},\alpha}$ & Charge operator of edge $\alpha$ within the QPC 
 & Eq.~\eqref{eq:q_qpc} \\ 
 \hline
 $\hat{\mathcal{I}}_\alpha$ & $\partial_t \hat{Q}_\alpha$ 
 & Eq.~\eqref{eq:charge_reduction} \\ 
 \hline
 $\mathcal{G}_{\alpha\alpha'}$ & Response of $\hat{\mathcal{I}}_\alpha$ to voltage bias in edge $\alpha'$ 
 & Eq.~\eqref{eq:response_definition} \\ 
 \hline
 $\mathcal{C}_{\alpha\alpha'}$ & Correlation between $\hat{\mathcal{I}}_\alpha$ and $\hat{\mathcal{I}}_{\alpha'}$ 
 & Eq.~\eqref{eq:smn} \\ 
 \hline
 $\hat{O}^{\text{eff}}$ & Operator $\hat{O}$ of a finite-size QPC 
 & Eq.~\eqref{eq:current_op_conservation} \\ 
 \hline
\end{tabular}
\caption{Summary of operators. The physical meanings and definitions of the operators used throughout this manuscript are summarized within.}
\end{center}
\end{table*}
\end{widetext}

\section{The system and the principle relations}
\label{sec:system}

\begin{figure}
    \centering
    \includegraphics[width=0.95\linewidth]{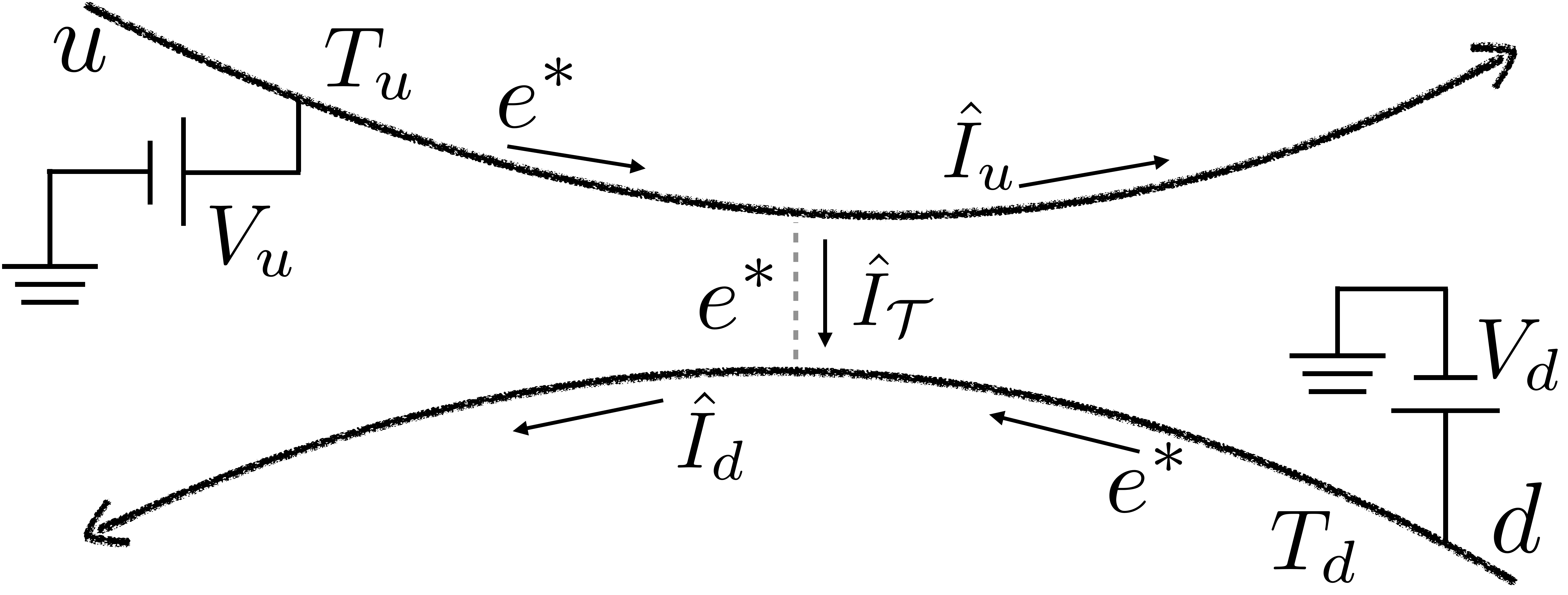}
    \caption{The system we consider in this work.
    The bulk of the system has a Laughlin filling fraction $\nu$.
    Two chiral Laughlin states, 1 and 2, each carrying quasiparticles with charge $e^*$, are thus hosted by the edge of the sample.
    At the upstream of the tunneling QPC, edges $u$ and $d$ are characterized by their own voltages ($V_u$ and $V_d$) and temperatures ($T_u$ and $T_d$), respectively.
    Two edges communicate through a tunneling QPC (located at $x = 0$, with quasiparticle $e^*$), with $\hat{I}_\mathcal{T}$ [cf. Eq.~\eqref{eq:current_op}] the corresponding tunneling-current operator.
    For later convenience, here we choose the convention that spatial coordinates increase along the transport direction, for both channels $u$ and $d$.
    Alternatively, in both channels, $x > 0$ at the downstream direction of the QPC, and $x < 0$ otherwise.
    Here $\hat{I}_u$ and $\hat{I}_d$ [cf. Eq.~\eqref{eq:charge_current}] refer to the current operators in channels $u$ and $d$, respectively, at both the upstream and downstream directions of the QPC.
    }
    \label{fig:model}
\end{figure}

In this section, we outline the system we consider, which is followed by a brief introduction of our major results.
In short, here we we visit the system, as displayed in Fig.~\ref{fig:model}, that consists of two chiral edges that communicate through a single QPC.
A more detailed introduction to our system and the corresponding Hamiltonian will be provided in Secs.~\ref{sec:system_1} and \ref{sec:momentum}, for the ideal case and the case with an extended QPC, respectively.
The major conclusions, i.e., the non-equilibrium fluctuation relations, are then introduced by Sec.~\ref{sec:result}, along with a corresponding brief discussion.

For the convenience of the readers, we begin by providing the major achievements to be presented and discussed in Sec.~\ref{sec:result}, 
i.e.,
\begin{equation}
\begin{aligned}
    S_\mathcal{T}(0) &= -S_{ud}(0) + k_B (T_u + T_d) G,\\
    S_\mathcal{T} (0) &= S_{\alpha \alpha}(0) + 2 k_B T_\alpha G - 4 G_0 k_B T_\alpha,
\end{aligned}
\label{eq:noise_conductance_relation_brief}
\end{equation}
as the relation between three types of fluctuations, i.e., the tunneling current noise $S_\mathcal{T}(0)$ [cf. \eqref{eq:tunneling_current_noise}, defined with the tunneling current $\hat{I}_\mathcal{T} $ of Eq.~\eqref{eq:iud_general} and Fig.~\ref{fig:model}], the cross correlation $S_{ud}(0)$ and the auto correlation $S_{\alpha \alpha}$ [cf. Eq.~\eqref{eq:auto_cross_correlations} for $\alpha \neq \alpha'$ and $\alpha = \alpha'$, respectively, with current operators $\hat{I}_u$ and $\hat{I}_d$ defined by Eq.~\eqref{eq:charge_current}].
These two equations further involve the thermal fluctuation of channel $\alpha$, quantified by $k_B T_\alpha$, the conductance $G$ at the tunneling QPC [$\partial I_\mathcal{T}/\partial V$, cf. Eq.~\eqref{eq:noise_conductance_relation}], and the quantized charge conductance, $G_0 \equiv \nu e^2/h$.

As will be further discussed after Eq. Eq.~\eqref{eq:noise_conductance_relation}, the major results of Eq.~\eqref{eq:noise_conductance_relation_brief} are exactly valid in stationary systems hosting two chiral channels that are \textit{ab initio} in equilibrium (cf. Fig.~\ref{fig:model}), i.e., \textit{each} characterized by the corresponding stationary voltage and temperature before the tunneling area.
Within this area, the inter-edge communication is characterized by a time-independent tunneling amplitude, within the effective QPC-tunneling model~\cite{Fendley:1995}.
Importantly, Eq.~\eqref{eq:noise_conductance_relation_brief} applies to situations where $T_u \neq T_d$, when the tunneling at the QPC depends on the applied voltage, and cases with a spatially extended QPC or momentum-dependent tunneling amplitudes, when the tunneling QPC falls into categories discussed in Appendix~\ref{app:momentum_dependent}. 

As the requirements, the size of the extended QPC should remain much smaller than the charge and thermal equilibration length (considering different reservoir temperatures) such that it is legitimate to characterize particles belonging to the same edge by the same bias and temperature. Otherwise the energy scales governing the tunneling may differ from the biases and temperatures of reservoirs.
In addition, no source or drain should be contained by the extended QPC.
Notably, due to the chiral feature of the involved channels, the tunneling QPC is not necessarily limited to the weak-tunneling limit. This versatility regarding the experimentally relevant factors above distinguishes our work from the existing literature.

\subsection{The system}
\label{sec:system_1}

In this work, we consider a system at Laughlin filling fraction $\nu = 1/(1 + 2n)$ in the bulk of the sample, with $n$ as a positive integer.
The corresponding edge modes (cf. Fig.~\ref{fig:model}) can be considered as two chiral counterpropagating channels with index  $\alpha =u, d$.
Tunneling between them is enabled by a QPC, which allows the transport of quasiparticles carrying charges equal to integer multiples of $e^*$~\cite{PrangeGirvinBook} (ideally, $e^* = \nu e$).
The system is described by the full Hamiltonian
\begin{equation}
    H = H_0 + H_\text{bias} + H_\mathcal{T},
    \label{eq:full_hamiltonian}
\end{equation}
where
\begin{equation}
\begin{aligned}
    H_0 & = H_u + H_d = \frac{v}{4\pi} \sum_{\alpha =u,d} \int_{-\mathcal{L}}^\mathcal{L} \:\! d x \: [ \partial_x \phi_\alpha ( x )]^2 ,
\end{aligned}
\label{eq:lead_hamiltonian}
\end{equation}
which refers to the Hamiltonian for these two free chiral Laughlin channels.
Here $v$ is the velocity, and $\phi_\alpha$ corresponds to the bosonic mode of the channel $\alpha =u, d$,
 which obeys the bosonic commutation relation $[\phi_\alpha(x), \phi_{\alpha'} (x')] = i\pi\delta_{\alpha,\alpha'} \text{sgn} (x-x')$, with $\delta_{\alpha,\alpha'}$ as the Kronecker Delta, and $\text{sgn} (x-x')$ as the sign function, featuring channel chiralities.
Importantly, although these two channels of Fig.~\ref{fig:model} have different chiralities, we can nevertheless choose the convention such that $x$ increases along the transport direction of each channel.
For simplicity, here the system size $\mathcal{L} $ is assumed to be much larger than any other relevant length scales, such that it can be taken as infinity within most parts of this work.
These bosonic modes are related to anyonic operators in the corresponding channel, following the bosonization
\begin{equation}
    \psi_\alpha (x) = \frac{1}{\sqrt{2\pi l_c}} e^{i\sqrt{\nu} \phi_\alpha} (x),
    \label{eq:bosonization}
\end{equation}
at position $x$, with $l_c$ a short-distance cutoff.
In Eq.~\eqref{eq:bosonization}, we have neglected the Klein factors as they are not relevant to the topic of the current work (see, however, e.g., Refs.~\cite{Safi:2001,Guyon:2002,Kane:2003}, and discussions after Eq.~\eqref{eq:klein_factors} of Appendix~\ref{app:finite_size_dyson}).

As the second factor, these two Laughlin edges have different biases $V_u$ and $V_d$ (cf. Fig.~\ref{fig:model}), respectively, with $V \equiv V_u - V_d$ representing the bias difference, leading to the corresponding Hamiltonian
\begin{equation}
\begin{aligned}
    H_\text{bias} 
    & = e V_u \frac{\sqrt{\nu}}{2\pi} \int_{-\mathcal{L}}^\mathcal{L} dx \ \partial_x \phi_u + e V_d \frac{\sqrt{\nu}}{2\pi} \int_{-\mathcal{L}}^\mathcal{L} dx \ \partial_x \phi_d\\
    & =  V_u \hat{Q}_u + V_d \hat{Q}_d,
\end{aligned}
    \label{eq:hbias}
\end{equation}
where $\hat{Q}_\alpha$ refers to the total charge of edge $\alpha$, i.e.,
\begin{equation}
\begin{aligned}
    \hat{Q}_u & = \frac{\sqrt{\nu}}{2\pi} \left[ \phi_u (\mathcal{L}) - \phi_u (-\mathcal{L}) \right],\\
    \hat{Q}_d & = \frac{\sqrt{\nu}}{2\pi} \left[ \phi_d (\mathcal{L}) - \phi_d (-\mathcal{L}) \right],
\end{aligned}
\label{eq:charge_operator}
\end{equation}
with $2\mathcal{L}$ the size of both chiral Luttinger liquid edges, and the QPC located at their centers ($x=0$).
In general, $V_u$ and $V_d$ of Eq.~\eqref{eq:hbias} can be time-dependent. Here, we focus on the DC limit, where both are constant.
Crucially, Eq.~\eqref{eq:hbias} remains valid along the entire edge, though the chemical potential at the downstream of the tunneling area is ill-defined.
Actually, after the tunneling QPC, the quasiparticle density can be considered as the summation of a well-defined ``quasiparticle surface'' (e.g., fermi sea of a fermionic system), before the QPC, and these that tunnel at the QPC.
This change in charge density can be captured by adjusting $\partial_x \phi_\alpha$ in Eq.~\eqref{eq:hbias}~\footnote{As an example, in the limiting case of a fully transparent tunneling QPC, the two edges exchange their biases downstream of the QPC, as all particles with energies between the two chemical potentials are transferred from one channel to the other.}.

Free-edges $u$ and $d$, depicted by Eqs.~\eqref{eq:lead_hamiltonian} and \eqref{eq:hbias}, are connected via the tunneling area (cf. the dashed line of Fig.~\ref{fig:model}) that is characterized by the tunneling Hamiltonian $H_\mathcal{T}$, with the subscript $\mathcal{T}$ referring to the transfer between two edges.
The induced charge transfer can be further characterized by the tunneling charge current, $\hat{I}_\mathcal{T}$,
\begin{equation}
\begin{aligned}
 \hat{I}_\mathcal{T} & = -\frac{\partial}{\partial t}\hat{Q}_u = -\sqrt{\nu} \frac{\partial H_\mathcal{T}}{\partial\phi_u} \\
 & = \frac{\partial}{\partial t}\hat{Q}_d = \sqrt{\nu} \frac{\partial H_\mathcal{T}}{\partial\phi_d} ,
 \end{aligned}
 \label{eq:iud_general}
\end{equation}
in terms of charge operators $\hat{Q}_u$ and $\hat{Q}_d$, cf. Eq.~\eqref{eq:charge_operator}.
Importantly, Eq.~\eqref{eq:iud_general} remains strictly valid without specifying $H_\mathcal{T}$ (cf. Appendix~\ref{eq:iud_derivation} for the derivation).
This general applicability is crucial for the strict validity of our central result, Eq.~\eqref{eq:noise_conductance_relation_brief}, beyond the weak-tunneling limit and in the presence of practical complications such as an extended tunneling area, interactions therein, and a momentum-dependent tunneling amplitude.

Despite the inter-edge communication within the tunneling area, the current operators of each edge can be expressed in a uniform (i.e., both before or after the QPC) way, i.e.,
\begin{equation}
\begin{aligned}
\hat{I}_u (x) &\equiv \frac{e \sqrt{\nu}}{2 \pi }\, \partial_t \phi_u (x,t) ,  \\
\hat{I}_d (x)&\equiv  \frac{e \sqrt{\nu}}{2 \pi }\, \partial_t \phi_d (x,t) ,
\:
\end{aligned}
\label{eq:charge_current}
\end{equation}
in terms of bosonic operators $\phi_u$ and $\phi_d$, respectively.
Notice that in literatures e.g., Ref.~\cite{Fendley:1995c}, current operators are defined after differentiation over space [i.e., $\sim \partial_x \phi_u$ or $\partial_x \phi_d$, instead of that in Eq.~\eqref{eq:charge_current}].
Indeed, within the interaction picture, the differentiation over space can be transformed into that over time for systems containing chiral fields.
Within this work, we stick to Eq.~\eqref{eq:charge_current},
which is more convenient in the derivation of fluctuation relations (cf. Appendix~\ref{app:intermediate_finite_size}).

For later convenience, we can further define current averages (as a reminder, the QPC is located at $x = 0$),
\begin{equation}
\begin{aligned}
    &I_{u0} = \big\langle \hat{I}_u (x<0) \big\rangle, \ I_{d0} = \big\langle \hat{I}_d (x<0) \big\rangle,\\
    &I_{u} = \big\langle \hat{I}_u (x>0) \big\rangle, \ I_{d} = \big\langle \hat{I}_d (x>0) \big\rangle,\\
    & I_\mathcal{T} = \big\langle \hat{I}_\mathcal{T} \big\rangle,
\end{aligned}
\label{eq:current_averages}
\end{equation}
by evaluating the averages of current operators with respect to the full Hamiltonian.
Here $x < 0 $ and $x>0$ refer to the locations in the upstream and downstream directions, respectively, of the QPC.

For a finite-size tunneling area,
due to current conservation, current expectations satisfy $I_u (x_{u,\text{f}}) - I_u (x_{u,\text{i}}) + I_\mathcal{T} = 0$, where $x_{u,\text{i}}$ and $x_{u,\text{f}}$ refer to the boundaries of the tunneling area, at the upstream and downstream directions, respectively.
This conservation actually can be strictly proven, via Eq.~\eqref{eq:iud_general} and the equation of motion method within the Schrodinger picture. For instance, for edge $u$,
\begin{equation}
\begin{aligned}
    \hat{I}_\mathcal{T} & = -\frac{\partial}{\partial t}\hat{Q}_u = - \frac{\partial}{\partial t}\hat{Q}_\text{QPC,u}\\
    & = - \frac{\sqrt{\nu } e}{2 \pi }\frac{\partial}{\partial t}\left[ \phi_u (x_{u,\text{f}}) - \phi_u (x_{u,\text{i}}) \right]\\
    & = - \hat{I}_u (x_{u,\text{f}}) + \hat{I}_u (x_{u,\text{i}}),
\end{aligned}
\label{eq:current_op_conservation}
\end{equation}
where
\begin{equation}
    \hat{Q}_\text{QPC,u} \equiv \frac{\sqrt{\nu} e}{2\pi} \left[ \phi_u (x_{u,\text{f}}) - \phi_u (x_{u,\text{i}}) \right],
    \label{eq:q_qpc}
\end{equation}
refers to the charge operator for the tunneling area of edge $u$, i.e., between $x_{u,\text{i}}$ and $x_{u,\text{f}}$.
For an ideal QPC, $x_{u,\text{i}}$ and $x_{u,\text{f}}$ approach the tunneling point ($x = 0$) from the upstream and downstream directions, respectively.
Eq.~\eqref{eq:current_op_conservation} has taken the assumption that edges $u$ and $d$ exchange quasiparticles only within the tunneling area.
In addition, both edges are assumed to stay stationary, such that current operators at different reservoir-system boundaries cancel out.

Though the explicit expression of $H_\mathcal{T}$ does not enter either the current operators ($\hat{I}_\mathcal{T}$, $\hat{I}_u$ and $\hat{I}_d$) or the derivation of the main conclusion [i.e., Eq.~\eqref{eq:noise_conductance_relation_brief}, cf. Appendix~\ref{app:intermediate_finite_size} for details of derivation], here, for the convenience of the readers, we provide the general expression of $H_\mathcal{T}$, as,
\begin{equation}
\begin{aligned}
    H_\mathcal{T} & = \sum_{n \ge 1} (A_n + A_n^\dagger), \\
    A_n & = \frac{\xi_n}{2\pi l_c} \exp \left\{-i n  \sqrt{\nu}\left[ \phi_u (0) - \phi_d (0 ) \right] \right\}, \\
\end{aligned}
    \label{eq:ht}
\end{equation}
where the argument $x = 0$ refers to the position of the QPC, and $\xi_n$ is the tunneling amplitude of the corresponding high-order tunneling operator $A_n$ that transport quasiparticles with charge $ne^*$ between two edges.
Importantly, we would like to stress that $A_n$ is not the only operator that can introduce the $n$th order tunneling at the collider.
Indeed, Eq.~\eqref{eq:ht} with $n=1$ already suffices to generate higher-order tunneling processes, including the bunching of anyons, via Keldysh expansion when evaluating correlation functions.
Actually, following Ref.~\cite{Fendley:1995c}, when the tunneling between two chiral Laughlin channels is faithfully described by the $n=1$ term of Eq.~\eqref{eq:ht}, the quantum transport is integrable following the thermodynamic Bethe ansatz, for not only the weak-tunneling limit, but also the strong-tunneling limit and the in-between crossover regime, where higher-order tunnelings are necessarily included by the latter two.
However, the amplitude of a higher-order term generated by the leading ($n=1$) operator is functions of $\xi_1$, i.e., $\xi_n \propto \xi_1^n$.
We thus include $n\neq 1$ terms in Eq.~\eqref{eq:ht}, such that $\xi_n$ can be more generally treated as independent.

With the tunneling Hamiltonian of Eq.~\eqref{eq:ht},
the corresponding tunneling current operator then becomes
\begin{equation}
\begin{aligned}
    \hat{I}_\mathcal{T} & = -\frac{\partial}{\partial t}\hat{Q}_u = \frac{\partial}{\partial t}\hat{Q}_d \\
    & = - e^* \frac{1}{\pi l_c  }\: \sum_{n\ge 1} n\  \xi_n \sin \left\{ n\sqrt{\nu}\left[ \phi_u (0) - \phi_d (0 ) \right] \right\}\\
    & = i e^* \sum_{n\ge 1} n \left( A_n^\dagger - A_n \right),
\end{aligned}
\label{eq:current_op}
\end{equation}
where the prefactor $e^*$ equals the fractional charge of a single quasiparticle carried by a Laughlin edge with a filling fraction $\nu$.
In both Eqs.~\eqref{eq:ht} and \eqref{eq:current_op}, the tunneling amplitudes $\xi_n$ depend on the details of the tunneling area.
These details, as has been stated, however do not influence our main conclusion of Eq.~\eqref{eq:noise_conductance_relation_brief}.

\subsection{Momentum-dependent transmission amplitude}
\label{sec:momentum}

In Eqs.~\eqref{eq:ht} and \eqref{eq:current_op}, we focus on the ideal situation where the transmission occurs at a single point, with a constant transmission amplitude.
In practice, Eq.~\eqref{eq:current_op} is however not necessarily valid, due to e.g., an extended quantum point contact~\cite{DolcettoSassettiPRB12}, the presence of an effective dot in the colliding area~\cite{KaneFisherPRB92b,Kane:1992,MatveevLarkinPRB92,FurusakiNagaosaPRB93,TakeiMilletariRosenowPRB10,Mebrahtu12,DongHuaixiuPRB14}, or systems under the influence of various (e.g.,electron-electron and or electron-phonon) interactions~\cite{RosenowHalperinPRL02,PapaMacDonaldPRL04,CarregaSassettiPRL11,SnizhkoCheianovPRB15}.
For the convenience of the readers, we explicit show the general influence of these factors on the tunneling Hamiltonian and the tunneling current operator (cf. more discussions in Appendix~\ref{app:momentum_dependent}).
Before moving on, we nevertheless stress that our major result, i.e., Eq.~\eqref{eq:noise_conductance_relation_brief}, actually does not require the explicit expression of the tunneling current operator.
Instead, the definition Eq.~\eqref{eq:iud_general}, which is actually used in the prove of Eq.~\eqref{eq:noise_conductance_relation_brief}, remains valid even under the influence of factors mentioned above.

As the starting point, with influencing factors mentioned above,
the tunneling Hamiltonian can be effectively written into a more general form (cf. Appendix~\ref{app:momentum_dependent} for the discussion of several examples),
\begin{equation}
\begin{aligned}
    &H_\mathcal{T}^\text{eff} 
     =  \sum_{n\ge 1} \left[ A_n^\text{eff} + \left(A_n^\text{eff}\right)^\dagger \right] \\
    & = \iint  dx \ dx' \sum_n \xi_n^\text{eff} (x,x') e^{-in\sqrt{\nu} \left[ \phi_u(x) - \phi_d(x') \right]} + h.c.,\\
    & A_n^\text{eff} = \iint dx\  dx' \xi_n^\text{eff} (x,x') e^{-i n \sqrt{\nu} [\phi_u(x) - \phi_d(x')] } ,
\end{aligned}
\label{eq:ht_eff}
\end{equation}
with $\xi_n^\text{eff} (x,x')$ the tunneling amplitude from position $x$ to $x'$ (of edges $u$ and $d$, respectively), of the $n$th-order tunneling operator $A_n^\text{eff}$.
Here the superscript ``eff'' stresses the fact that $H_\mathcal{T}^\text{eff}$ applies not only to a finite-size QPC, but can arise from momentum-dependent transmission at the QPC.
Indeed, due to the position-dependence of $\xi_n^\text{eff} (x,x')$, its Fourier-transformed counterpart, i.e.,
\begin{equation}
    \xi_{n;k,k'}^\text{eff} \equiv \iint dx dx' e^{ikx - ik'x'} \xi_n^\text{eff} (x,x'),
\end{equation}
becomes momentum dependent.
This dependence indicates that with the extended QPC, the tunneling barrier has different influences on
quasiparticles with different momentum (and hence different energies).
On the contrary, $\xi_{n;k,k'}^\text{eff} $ is simply a constant number for an ideal QPC, $\xi_n^\text{eff} (x,x')\sim \delta (x-x')$, where tunneling occurs at a single point.
Corresponding to the effective tunneling Hamiltonian Eq.~\eqref{eq:ht_eff}, the effective tunneling operator can be written as:
\begin{equation}
    \hat{I}_\mathcal{T}^\text{eff} = i \sum_{n\ge 1} n e^* \left[ - A_n^\text{eff} + \left(A_n^\text{eff}\right)^\dagger \right] ,
    \label{eq:it_eff}
\end{equation}
after including tunneling-current contributions from all positions within the finite-size tunneling area.
As the reminder, Eq.~\eqref{eq:iud_general}, which enters the derivation of Eq.~\eqref{eq:noise_conductance_relation_brief} in Appendix~\ref{app:intermediate_finite_size}, remains valid for the system with a finite-size QPC.

The dependence of the transmission amplitude on momentum is actually present in most experiments, except for recent reports of Ref.~\cite{VeillonNature24}.
We stress again that in Eq.~\eqref{eq:ht_eff}, the tunneling amplitude $\xi_n^\text{eff} (x,x')$ can describe not only a finite-size tunneling QPC. It can also effectively describe systems where tunneling at the QPC couples to other degrees of freedom, including e.g., electron-phonon or electron-electron interactions, or QPC-involved tunnelings.
For a better illustration, more relevant discussions will be provided in the Appendix~\ref{app:momentum_dependent}.
The voltage-dependent transmission amplitude, which is considered by Ref.~\cite{kyrylo_FQHE_2024}, is however beyond the scope of Eq.~\eqref{eq:ht_eff}.
The latter dependence will be discussed separately in Sec.~\ref{sec:voltage_dependence}.

Before proceeding, we would like to stress that with current operators provided in this section,
our main result, i.e., Eq.~\eqref{eq:noise_conductance_relation_brief}, applies to systems hosting a finite-size tunneling QPC.
Interestingly, Eq.~\eqref{eq:iud_general}, as an alternative definition of the tunneling current operator, also remains valid for an extended QPC.

\subsection{Definition of noises and the major results}
\label{sec:result}

With current operators introduced above, the system noise can be described by the tunneling current, and three correlation functions. The latter functions include the tunneling current noise, the cross correlation, and the auto correlation.

As the starter, we can define the finite-frequency tunneling current noise as the fluctuation of the tunneling current, i.e.,
\begin{equation}
\begin{aligned}
    S_\mathcal{T} (\omega) & \equiv 2 \int_{-\infty}^\infty dt e^{i\omega t} \Big\langle  \delta \hat{I}_\mathcal{T} (t) \delta \hat{I}_\mathcal{T} (0)  \Big\rangle,
\end{aligned}
\label{eq:tunneling_current_noise}
\end{equation}
for the interested frequency $\omega$, 
and $\delta\hat{I}_\mathcal{T} \equiv \hat{I}_\mathcal{T} - I_\mathcal{T} $ is the operator referring to the fluctuation of tunneling current.
Here the correlation is evaluated with respect to the full Hamiltonian.
Notice that in Eq.~\eqref{eq:tunneling_current_noise}, the definition of $S_\mathcal{T}(\omega)$ differs from that of Refs.~\cite{Campagnano:2016,OneHalfPRB24}, both due to the extra factor of two in front and the missing symmetrization (the anti-commutator).
We stress, however, that when $\omega \to 0$, which is the focused situation of our work, these two definitions are actually equivalent to each other.

Alternatively, the system noise can also be quantified by the auto- and cross correlations, i.e.,
\begin{equation}
\begin{aligned}
S_{\alpha \alpha'} (x_\alpha,x_{\alpha'}; \omega) & \equiv 2\int_{-\infty}^\infty dt e^{i\omega t} \big\langle \delta\hat{I}_\alpha (x_\alpha, t) \delta\hat{I}_{\alpha'} (x_{\alpha'}, 0)  \big\rangle,
\end{aligned}
\label{eq:auto_cross_correlations}
\end{equation}
where $\delta \hat{I}_\alpha \equiv \hat{I}_\alpha - I_\alpha $ refers to the fluctuation of current in channel $\alpha = u,d,$ downstream of the QPC.
Like Eq.~\eqref{eq:tunneling_current_noise} for the tunneling current noise $S_\mathcal{T}$, correlation function of Eq.~\eqref{eq:auto_cross_correlations} is presented without symmetrization.
Eq.~\eqref{eq:auto_cross_correlations} refers to the autocorrelation of the current in the same channel, when $\alpha = \alpha'$; When $\alpha \neq \alpha'$, it instead refers to the cross correlation between currents in different channels, after scattering at the QPC.
Practically, for negligible equilibration, Coulomb interaction, and heat diffusion, currents $I_u$ and $I_d$ and the associated noises of a chiral system can be experimentally measured at positions that are distant from the tunneling QPC.
The experimental measurements of the auto and cross correlations are thus more accessible in comparison to the direct measurement of the tunneling current noise. Indeed, the latter requires a complicated local measurement at the QPC that can be potentially affected by, e.g., crosstalk among local gates and interactions between edge modes.

Among our central results, the tunneling current noise and two current-current correlation functions are connected via the non-equilibrium fluctuation relation [as a reminder, $V\equiv V_u-V_d$ in $H_\text{bias}$, cf. Eq.~\eqref{eq:hbias}]
\begin{equation}
\begin{aligned}
    S_\mathcal{T}(0) &= -S_{ud}(0) + k_B (T_u + T_d) \frac{\partial I_\mathcal{T}}{\partial V},\\
    S_\mathcal{T}(0) &= S_{\alpha \alpha}(0) + 2 k_B T_\alpha \frac{\partial I_\mathcal{T}}{\partial V}  - 4 G_0 k_B T_\alpha,
\end{aligned}
\label{eq:noise_conductance_relation}
\end{equation}
where $k_B$ is the Boltzmann constant, and the argument ``$(0)$'' indicates the zero-frequency noise. 
The first line of Eq.~\eqref{eq:noise_conductance_relation} contains the conductance that is defined with the tunneling current average $I_\mathcal{T}$ [cf. Eq.~\eqref{eq:current_averages}].
Again $\alpha  =u, d$ refers to two autocorrelation functions, $G_0 \equiv \nu e^2/h$ refers to the ``quantized'' conductance of a single chiral Laughlin edge with filling fraction $\nu$, and $T_u$, $T_d$ are the temperatures of channels $u$ and $d$, respectively, at the \textit{upstream} of the QPC.
We stress that in systems under current consideration (i.e., that of Fig.~\ref{fig:model}, containing two chiral channels that communicate through the QPC), Eq.~\eqref{eq:noise_conductance_relation} is exact, i.e., being valid for all values of the bias difference $V$ and the tunneling probability $\xi$ through the central QPC [cf. Eq.~\eqref{eq:ht}]. In addition, it tolerates a difference between the temperatures $T_u$ and $T_d$, such that the relations of Eq.~\eqref{eq:noise_conductance_relation} perfectly apply to the research on delta-T noise, i.e., the noise induced by a bias in temperature.
More importantly, Eq.~\eqref{eq:noise_conductance_relation} allows for an momentum dependency of the transmission probability (cf. Appendix~\ref{app:momentum_dependent}), which are, importantly, crucial realistic factors in practical experiments.
Its modification in systems with a bias-dependent transmission probability is discussed later in Sec.~\ref{sec:voltage_dependence}.
These latter three improvements, sharply distinguishing our work from previous literatures~\cite{Kane:1994b,Fendley:1996,WangFeldmanPRL13,FeldmanMotyPRB17,OneHalfPRB24}, are arrived due to our technical advances in dealing with bosonized out-of-equilibrium backscattering models (cf. Appendix~\ref{app:finite_size_dyson}).
Nevertheless, we would like to stress that the simple zero-frequency relations emphasized in Eq.~\eqref{eq:noise_conductance_relation} are specialized to stationary reservoir-driven Laughlin edges, with incoming channels characterized by voltages and temperatures. They should not be read as relations for arbitrary non-thermal (or Floquet) incoming edge states.

Notice that the second line of Eq.~\eqref{eq:noise_conductance_relation}, concerning auto correlation and the tunneling current noise has been reported by e.g., Refs.~\cite{Kane:1994b,Fendley:1996,FeldmanMotyPRB17}, however without discussions on its validity in realistic experimental setups.
The first line, concerning cross correlation and the tunneling current noise is instead strictly visited by Ref.~\cite{OneHalfPRB24}, however only for a $\nu = 1/2$ system via refermionization.
Our current work goes beyond these previous studies from two major perspectives.
Firstly, when considering the relation between cross correlation and the tunneling current noise, our work goes beyond Ref.~\cite{OneHalfPRB24}, which is limited to only $\nu = 1/2$, to Laughlin system with general filling fractions $\nu$.
It is further shown explicitly that for both auto and cross correlations, the non-equilibrium fluctuation relation Eq.~\eqref{eq:noise_conductance_relation} can be obtained within the same framework~\cite{DolciniPRB05,BenaSafiPBR07,ines_proceedings_2009,ines_philippe}, which is actually beyond linear response, cf. Sec.~\ref{sec:generalized_linear_response} for more details.
Secondly, our work considers two experimentally practical factors that are neglected by Refs.~\cite{Kane:1994b,Fendley:1996,FeldmanMotyPRB17,OneHalfPRB24}: (i) A temperature inequality $T_u \neq T_d$, and (ii) The momentum dependence of the transmission amplitude [cf. Eq.~\eqref{eq:ht_eff} of Sec.~\ref{sec:momentum}, and Appendix~\ref{app:momentum_dependent} for more detailed discussions].
Actually, the validity of Eq.~\eqref{eq:noise_conductance_relation} is proven under these two experimentally practical factors above.
We further show that
Eq.~\eqref{eq:noise_conductance_relation} can be extended to systems with a bias-dependent $\xi$ (cf. discussions in Sec.~\ref{sec:voltage_dependence}), given modifications introduced.
Considering the frequent presence of these factors in real experiments, our work greatly supports the validity of Eq.~\eqref{eq:noise_conductance_relation} in practical experiments.

\begin{figure}
    \centering
    \includegraphics[width=0.99\linewidth]{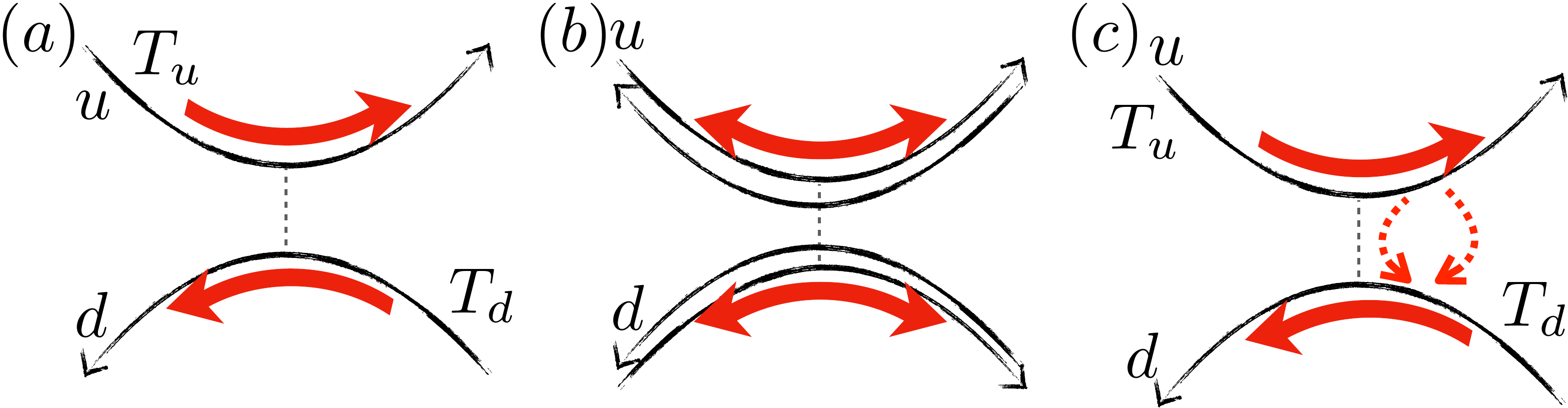}
    \caption{Heat flow of different structures.
    Here the black arrows highlight the flow of quasiparticles in each channel.
    The red arrows, being thick and dashed, instead refer to the heat flow carried by electrons and phonons, respectively.
    (a) The currently considered setup, which consists of two chiral channels. We assume a weak electron-phonon coupling, such that heat current flows along the same direction of the charge current.
    Of this case, the channel temperatures are well defined at the upstream of the QPC (indicated by the dashed line).
    (b) and (c) Setups containing non-chiral fermionic channels, or a strong phonon-induced heat diffusion, respectively.
    Of either case, edge temperatures at the upstream of the tunneling QPC become ill-defined.
    Especially, in (c), $T_d$ becomes ill-defined (even at the upstream of the QPC) after receiving the heat flow from channel $u$, through the phonon-induced heat diffusion (indicated by the red dashed arrows).
    }
    \label{fig:heat_flow}
\end{figure}

Before proceeding further, here we pause a while to briefly discuss the validity of Eq.~\eqref{eq:noise_conductance_relation}, when two temperatures are different.
Firstly, here we choose to use the temperatures at the upstream of the QPC. This option is chosen since edges become non-equilibrium after tunneling at the QPC, making the definition of edge temperatures a complicated task (cf. Refs.~\cite{EngquistAndersonPRB81,LandscapePRL25} for the definition of an effective temperature for non-equilibrium systems).
Secondly, in obtaining Eq.~\eqref{eq:noise_conductance_relation}, we are using field theory with real-time argument (cf. Appendix~\ref{app:intermediate_finite_size}), where the temperature of each corresponding channel only enters the QPC-free part of the action [cf. $\mathcal{S}_u^0$ and $\mathcal{S}_d^0$ of Eq.~\eqref{eq:generating_functional}].
The other part of the action [cf. $\mathcal{S}_\mathcal{T}^0$ of Eq.~\eqref{eq:generating_functional}] that models the tunneling between two edges thus becomes explicitly temperature independent.
This is an important observation since the temperature
at the QPC is actually undetermined when two involved edges have different temperatures.
Finally, we stress that even at the upstream of the QPC, the temperature $T_u$ and $T_d$ are only well defined, given the satisfactory of two requirements [cf. Fig.~\ref{fig:heat_flow}(a)]: (i) the chiral feature of both channels, (ii) a negligible, if any, phonon-induced heat diffusion.
More pedagogically, the size of the scattering area should be much smaller than the equilibration length, but much larger than the thermal diffusion length.
The violation of either requirement actually leads to the ill-definition of the temperature of involved channels.

Indeed, in the setup of Fig.~\ref{fig:heat_flow}(b), when edge states are non-chiral, the heat that tunnels through the QPC has the chance, via the mode with the opposite chirality, to return to the QPC, leading to ill-defined channel temperatures at the upstream of the QPC.
Alternatively, assuming $T_u > T_d$, a strong enough heat diffusion (indicated by the red dashed arrows) in the setup of Fig.~\ref{fig:heat_flow}(c) can change the temperature of channel $d$ at the upstream of the QPC, to deviate from its source value, $T_d$.
Like that of Fig.~\ref{fig:heat_flow}(b), here the temperatures of both edges at the upstream of the tunneling QPC are ill-defined.
More specifically, in-practice the characteristic length of the heat diffusion is requested to be much smaller than the size of the QPC.

Finally, for the setup of Fig.~\ref{fig:heat_flow}(a), Eq.~\eqref{eq:noise_conductance_relation} remains generically valid even in the presence of a short-range Coulomb interaction.
Firstly, when this range of interaction is smaller than the size of the QPC, the influence of Coulomb interaction is simply negligible.
More interestingly, when the range of interaction becomes larger than the size of QPC, the corresponding problem becomes effectively equivalent to the situation with an extended colliding area, with the effective tunneling Hamiltonian Eq.~\eqref{eq:ht_eff}, cf. Appendix~\ref{app:momentum_dependent}.

Before the closure of this section, we notice that in Eq.~\eqref{eq:noise_conductance_relation}, the filling fraction explicitly influences only the conductance $G_0$ of a single chiral channel.
The other terms are however not explicitly influenced by the filling fraction or the scaling factor of the system --- the latter can in-principle become modified by system interactions.
As the consequence, in interaction-present systems, Eq.~\eqref{eq:noise_conductance_relation} remains valid (after accordingly redefining $G_0$), as long as the interaction does not sabotage or modify chiral features of both edges.

\section{Generalized out-of equilibrium linear response formulas for non-equilibrium systems}
\label{sec:generalized_linear_response}

Practically, an enormous amount of enlightening physical phenomena require the system to be out-of-equilibrium. They are thus clearly beyond the scope of conventional linear response, which requires the system to stay in equilibrium or close to equilibrium.
As a potential solution of non-equilibrium features, the framework of Refs.~\cite{DolciniPRB05,BenaSafiPBR07,ines_proceedings_2009,ines_philippe} is introduced to study the system response to an infinitesimal variation of the system drive, where the system itself can be (even strongly) out of equilibrium.
In this section, we present corresponding formulas that are relevant to our current study, following materials of Refs.~\cite{DolciniPRB05,BenaSafiPBR07,ines_proceedings_2009,ines_philippe}.
Before moving on, we mention that the extended Kubo formulas were derived in Ref.~\cite{FDT_nonlinear_kubo_Svirskii_JETP_1981} for systems under a finite DC voltage, but only within the framework of thermal states.
In addition, an experimental measurement has been achieved in a nonlinear quantum conductor~\cite{FDT_kubo_altimiras_2026}, following the theoretical proposal of Ref.~\cite{LesovikJETP1997}.

Rather than discussing Fig.~\ref{fig:model} that is mainly focused on by the current work, within this section we start with the more general situation, where $\alpha$ equals 1 to $N$, with $N$ the total number of involved edges.
Among the central messages of the non-equilibrium response theory for multiterminal nonlinear systems~\cite{SafiBenaCrepieuxPRB08,ines_proceedings_2009,ines_proceedings_2009,ines_philippe}, it is stated that (see Appendix~\ref{app:genearlized_linear_response} for more details)
\begin{equation}
    \frac{\partial \mathcal{I}_\alpha (t)}{\partial V_{\alpha'} (t')} 
    = -\frac{i}{\hbar} \Theta (t-t') \Bigg\langle \left[ \hat{\mathcal{I}}_\alpha (t),  \frac{\partial H (t')}{\partial V_{\alpha'} (t')} \right] \Bigg\rangle,
\label{eq:interaction_eom}
\end{equation}
which gives the response of the current in edge $\alpha$, i.e., $\mathcal{I}_\alpha (t) \equiv \langle \hat{\mathcal{I}}_\alpha (t) \rangle$, to an infinitesimal modification of the voltage $V_{\alpha'}$ applied to channel $\alpha'$.
Crucially, here
\begin{equation}
    \hat{\mathcal{I}}_\alpha \equiv \partial_t \hat{Q}_\alpha
    \label{eq:charge_reduction}
\end{equation}
reflects the change (rate of enhancement) of the \textit{total} charge number $\hat{Q}_\alpha$ of reservoir $\alpha$.
It is thus in-principle different from $\hat{I}_\alpha (x)$ defined in Eq.~\eqref{eq:charge_current}, which is instead related to the \textit{local} operator.
Indeed, for our interested two-edge case, $\hat{\mathcal{I}}_u = - \hat{\mathcal{I}}_d$ is actually the tunneling current operator.
To avoid confusion with fluctuation operators, in Eq.~\eqref{eq:interaction_eom} we use $\partial$ to denote both ordinary and functional derivatives.
In addition, all correlation functions are evaluated with respect to the full Hamiltonian of the system, $H$.
Within our work, the voltage is assumed to couple linearly to the corresponding charge operator $\hat{Q}_{\alpha}$, cf. Eq.~\eqref{eq:hbias}.
Within this setup, and under the assumption of a bias-independent transmission amplitude $\xi$, Eq.~\eqref{eq:interaction_eom} further simplifies to
$\partial H / \partial V_{\alpha} = \hat{Q}_{\alpha}$.
The more general situation, involving a bias-dependent $\xi$, is addressed only in Sec.~\ref{sec:voltage_dependence}.
Other than the bias-independency of $\xi$,
derivations within this section actually apply to general expressions of the tunneling Hamiltonian.
From this perspective, Eq.~\eqref{eq:interaction_eom} remains valid, even when the tunneling Hamiltonian at the QPC has a momentum dependent transmission amplitude (see more details from Sec.~\ref{app:finite_size_dyson}), or if two edges have different temperatures.

We can further define the elements of this response function,
\begin{equation}
\begin{aligned}
    \mathcal{G}_{\alpha \alpha' } (t,t') \equiv & \frac{\partial \mathcal{I}_\alpha (t)}{\partial V_{\alpha'} (t')}  \\
    &= -\frac{i}{\hbar}\Theta (t-t') \big\langle [ \hat{\mathcal{I}}_\alpha (t), \hat{Q}_{\alpha'} (t') ] \big\rangle,
\end{aligned}
    \label{eq:response_definition}
\end{equation}
as the generalized charge conductance matrix elements $\mathcal{G}_{\alpha'} (t,t')$, in the dynamical (i.e., time-dependent) form.
Here in the second line we have taken $\partial H / \partial V_{\alpha} = \hat{Q}_{\alpha}$.
Following Eq.~\eqref{eq:response_definition}, and Eq.~\eqref{eq:iud_general} for the differentiation over charge operators, the differentiation of the generalized conductance over time equals
\begin{equation}
\begin{aligned}
    \frac{\partial \mathcal{G}_{\alpha \alpha'} (t,t')}{\partial t'} & = -\frac{i}{\hbar}\Theta (t-t')  \big\langle [ \hat{\mathcal{I}}_\alpha (t), \hat{\mathcal{I}}_{\alpha'} (t') ] \big\rangle\\
    & \ \ + \frac{i}{\hbar}\delta (t-t') \big\langle [ \hat{\mathcal{I}}_\alpha (t), \hat{Q}_{\alpha'} (t') ] \big\rangle,
\end{aligned}
\label{eq:g_differentiation}
\end{equation}
with which we arrive at the non-equilibrium fluctuation relation in the form of matrix elements, i.e.,
\begin{equation}
\begin{aligned}
    & i \hbar\left[ \frac{\partial \mathcal{G}_{\alpha \alpha'} (t,t')}{\partial t'} -  \frac{\partial\mathcal{G}_{\alpha' \alpha} (t',t)}{\partial t}\right]
    = \mathcal{C}_{\alpha \alpha'} (t,t') - \mathcal{C}_{\alpha' \alpha}(t',t),
\end{aligned}
\label{eq:noise_conductance_general}
\end{equation}
where we also define the elements of the noise matrix,
\begin{equation}
    \mathcal{C}_{\alpha \alpha'} (t,t') \equiv \big\langle \hat{\mathcal{I}}_{\alpha'} (t') \hat{\mathcal{I}}_{\alpha} (t) \big\rangle - \big\langle\hat{\mathcal{I}}_{\alpha'} (t') \big\rangle\big\langle \hat{\mathcal{I}}_{\alpha} (t) \big\rangle ,
    \label{eq:smn}
\end{equation}
as the correlation of current operators in channels $\alpha$ and $\alpha'$.
Notice that in obtaining Eq.~\eqref{eq:noise_conductance_general}, we are using the fact that for operators with the same argument in time (corresponding to the delta function) $[\hat{\mathcal{I}}_\alpha, \hat{Q}_{\alpha'}] \propto [ \hat{\mathcal{I}}_\alpha, \phi_{\alpha'} (\infty) ] - [ \hat{\mathcal{I}}_\alpha, \phi_{\alpha'} (-\infty) ] =0 $, as $\hat{\mathcal{I}}_\alpha$ actually involves only operators within the tunneling area.
With this observation, the second line of Eq.~\eqref{eq:g_differentiation} actually vanishes.

In systems with a time-independent Hamiltonian (which is actually our current situation), correlation functions $\mathcal{C}_{\alpha \alpha'}(t,t')$ and $\mathcal{G}_{\alpha \alpha'}(t,t')$ are both independent of the summation of moments, i.e., $t + t'$. Of this case, $\mathcal{C}_{\alpha \alpha'}(t,t') = \mathcal{C}_{\alpha \alpha'}(t - t')$ and $\mathcal{G}_{\alpha \alpha'}(t,t') = \mathcal{G}_{\alpha \alpha'}(t - t')$ such that we can define their Fourier transformed expressions as
\begin{equation}
\begin{aligned}
    \mathcal{G}_{\alpha \alpha'} (\omega)& \equiv \int d(t-t') e^{i\omega (t-t')} \mathcal{G}_{\alpha \alpha'}(t-t'),\\
    \mathcal{C}_{\alpha \alpha'} (\omega)& \equiv \int d(t-t') e^{i\omega (t-t')} \mathcal{C}_{\alpha \alpha'}(t-t').
\end{aligned}
\label{eq:fprim}
\end{equation}
With Fourier transformed functions of Eq.~\eqref{eq:fprim}, Eq.~\eqref{eq:noise_conductance_general} becomes alternatively
\begin{equation}
    \mathcal{C}_{\alpha \alpha'} (\omega) - \mathcal{C}_{\alpha' \alpha} (-\omega) = \hbar \omega \mathcal{G}_{\alpha \alpha'} (\omega) + \hbar \omega \mathcal{G}_{\alpha' \alpha} (-\omega).
    \label{eq:sg_relation}
\end{equation}
Eq.~\eqref{eq:sg_relation}, which can be considered as the generalized fluctuation-dissipation theorem,
is among the central conclusions obtained from Refs.~\cite{BenaSafiPBR07,ines_proceedings_2009,ines_philippe} in dealing with the bosonized out-of-equilibrium backscattering models.
It establishes the connection between conductances and the finite-frequency noise, that is valid for all voltage biases, temperatures, and transmission amplitudes through the QPC.
Remarkably, Eq.~\eqref{eq:sg_relation} is valid for most systems, as long as only the bias Hamiltonian $H_\text{bias}$ has a linear dependence on the bias.
It is thus highly applicable to practical experimental setups.
Notably, Eq.~\eqref{eq:sg_relation} has also been obtained by Refs.~\cite{GavishPRB00,Gavish04}, where this relation is addressed as non-equilibrium Kubo formula.
However, in these references, its applicability to quantum circuit remains unclear.

Before moving on, we provide the simplified version of the results discussed above, for our focused setup, i.e., that presented by Fig.~\ref{fig:model}.
As the major simplification, due to the existence of only two edges, $\hat{\mathcal{I}}_u = -\hat{\mathcal{I}}_d = -\hat{I}_\mathcal{T}$, so that while Eqs.~\eqref{eq:noise_conductance_general} and ~\eqref{eq:sg_relation} remain unchanged, the expression of the noise, given by Eq.~\eqref{eq:smn}, now becomes modified into
\begin{equation}
\begin{aligned}
    \mathcal{C}_{ud} (t,t')& = \big\langle \hat{\mathcal{I}}_d (t') \hat{\mathcal{I}}_u (t) \big\rangle - \big\langle\hat{\mathcal{I}}_d (t') \big\rangle\big\langle \hat{\mathcal{I}}_u (t) \big\rangle\\
    & = -\big\langle \hat{I}_\mathcal{T} (t') \hat{I}_\mathcal{T} (t) \big\rangle + \big\langle\hat{I}_\mathcal{T} (t') \big\rangle\big\langle \hat{I}_\mathcal{T} (t) \big\rangle.
\end{aligned}
\label{eq:two_edge_c}
\end{equation}
One can further remove the average product by introducing another function, i.e.,
\begin{equation}
\begin{aligned}
    & \left[ \mathcal{C}_{ud} (t,t') \!-\! \mathcal{C}_{ud} (t,t') \right]=\big\langle \hat{I}_\mathcal{T} (t) \hat{I}_\mathcal{T} (t') \big\rangle \!-\! \big\langle\hat{I}_\mathcal{T} (t') \hat{I}_\mathcal{T} (t) \big\rangle.
\end{aligned}
\label{eq:noise_difference}
\end{equation}
In addition, with this simplification, elements of the conductance matrix, e.g., $\mathcal{G}_{ud}$ defined by Eq.~\eqref{eq:response_definition} actually refers to the charge conductance, i.e.,
\begin{equation}
\begin{aligned}
    \lim_{\omega \to 0} \mathcal{G}_{ud} (\omega) & = -\frac{\partial I_\mathcal{T}}{\partial V_d} \Big|_\text{DC},\\
    \lim_{\omega \to 0} \mathcal{G}_{du} (\omega) & = \frac{\partial I_\mathcal{T}}{\partial V_u} \Big|_\text{DC},
\end{aligned}
\label{eq:dc_conductance}
\end{equation}
such that their summation basically produces the DC tunneling conductance through the tunneling area.

\section{Derivations on the non-equilibrium fluctuation relations}
\label{sec:noise_conductance_relation}

Following the previous section, especially the generalized fluctuation-dissipation theorem, Eq.~\eqref{eq:sg_relation}, we are ready to provide details in arriving at the major result of this work, i.e., the non-equilibrium fluctuation relation presented by Eq.~\eqref{eq:noise_conductance_relation}.

\subsection{QPC-free Green's functions and correlation functions}

As our strategy, we begin with the QPC-free situations [i.e., $\xi = 0$ in Eq.~\eqref{eq:ht}], where two disconnected edges, $u$ and $d$, are in-equilibrium by themselves.
These two channels are thus described by the equilibrium correlation functions~\cite{GiamarchiBook,KamenevBook,MahanBook} for non-interacting bosonic operators (with $v$ the velocity), i.e.,
\begin{equation}
\begin{aligned}
    c^\mathcal{A}_\alpha (x,x',\omega)
    & \!\! \equiv \!\!\!\int \!\! dt e^{i\omega t} \!\left[ -i\theta (-t) \big\langle [ \dot{\phi}_\alpha (x,t), \phi_\alpha (x',0) ] \big\rangle_0 \right]\\
    & = - 2\pi e^{- i \nu \omega (x - x')/v} \Theta  (x - x'),\\
    c^\mathcal{R}_\alpha (x,x',\omega)
    & \equiv \int dt e^{i\omega t} i \theta (t) \big\langle [ \dot{\phi}_\alpha (x,t), \phi_\alpha (x',0) ] \big\rangle_0 \\
    & = 2\pi e^{- i \nu \omega (x - x')/v} \Theta [- (x - x')],\\
    c^\mathcal{K}_\alpha (x,x',\omega) & \equiv \int dt e^{i\omega t} \big\langle [ \dot{\phi}_\alpha (x,t), \phi_\alpha (x',0) ] \big\rangle_0 \\
    & = 2\pi e^{- i \nu \omega (x - x')/v} \coth \left( \frac{\hbar\omega}{2k_B T_\alpha} \right),
\end{aligned}
\label{eq:free_c_functions_space}
\end{equation}
where superscripts $\mathcal{A}$, $\mathcal{R}$ and $\mathcal{K}$ refer to ``advanced'', ``retarded'' and ''Keldysh'' branches, respectively, of these correlation functions~\cite{KamenevBook}.
In Eq.~\eqref{eq:free_c_functions_space}, functions can easily be defined alternatively, in terms of bosonic Green's functions in the tunneling-free limit, cf. Eq.~\eqref{eq:impurity_free_gfs} of the appendix.
Here
the subscript ``0'' indicates to evaluate correlation functions for the QPC-free Hamiltonian, $H_0 + H_\text{bias}$.
Importantly, in Eq.~\eqref{eq:free_c_functions_space}, Green's functions of different edges differ only in the edge temperature $T_\alpha$ (at the upstream of the QPC) for the Keldysh sector.
Indeed, these bias of a single channel becomes meaningless for disconnected edges.
Here both the advanced and retarded branches contain a theta function, indicating the chiral feature of involved Laughlin edges.
Also notice that temperatures $T_u$ and $T_d$ will only influence the Keldysh branch of the Green's function.
In Eq.~\eqref{eq:free_c_functions_space}, the retarded and advanced Green's functions of channel $u$ equal that of channel $d$ (in contrast to that of Refs.~\cite{ines_proceedings_2009,ines_philippe}).
This is due to the fact that in both edges $u$ and $d$, the spatial coordinate is defined to increase along the transport direction.

Before further proceeding, we stress that QPC-free correlation functions Eq.~\eqref{eq:free_c_functions_space} are only valid for interaction-free channels.
Consequently, derivations that follow apply only to situations where two involved edges are interaction-free, though the more general dissipation-fluctuation relation, Eq.~\eqref{eq:sg_relation}, remains valid even for interacting systems.
Actually, as long as the chiral feature of both edges remains intact from interaction influences, 
these influences can be included by correspondingly modifying the anyonic signature, i.e., the fractional scaling dimension and fractional charge in 
Eq.~\eqref{eq:free_c_functions_space}.

Before moving to include influences of the QPC, here we provide the auto and cross correlation functions of the \textit{QPC-free} ($\xi=0$) case
\begin{equation}
\begin{aligned}
    s_{\alpha \alpha} & = \nu \frac{e^2 \omega}{\pi} \coth \left( \frac{\hbar\omega}{2k_B T_\alpha} \right),\\
    s_{ud} & = 0,
\end{aligned}
\label{eq:free_noises}
\end{equation}
where $s_{ud} = 0$, due to the missing of correlation when two channels are independent from each other.
The corresponding auto correlation $s_{\alpha \alpha}$ instead equals the Johnson–Nyquist noise, $4 G_0 k_B T_\alpha$, in the $\omega \to 0$ limit [cf. Eq.~\eqref{eq:f_c-term-final}], where $G_0 = \nu e^2/h$ is the conductance of an in-equilibrium chiral Luttinger liquid channel with filling fraction $\nu$.

\subsection{Influence of the tunneling QPC: relation between correlations and the tunneling current noise}

With the QPC introduced, tunneling between two edges brings in modifications on free Green's functions Eq.~\eqref{eq:free_greens_functions_space}.
For later convenience, here we define these modifications as $\delta S_{\alpha \alpha'} \equiv S_{\alpha \alpha'} - s_{\alpha \alpha'} $, for both the auto ($\alpha = \alpha'$) and cross ($\alpha \neq \alpha'$) correlations.
More specifically, following detailed derivations (cf. the Appendix~\ref{app:intermediate_finite_size}, based on Ref.~\cite{DolciniPRB05,BenaSafiPBR07,ines_proceedings_2009,ines_philippe}), we obtain explicit expressions of these modifications,
\begin{equation}
    \delta S_{\alpha \alpha'} (x,y, 0) = S^A_{\alpha \alpha'} (x,y, 0) + S^C_{\alpha \alpha'} (x,y, 0)
    \label{eq:noise_relation},
\end{equation}
where the third argument of these functions refers to the frequency, $\omega = 0$.
In Eq.~\eqref{eq:noise_relation}, the superscripts $A$ and $C$ highlight the involvement of anti-commutator and commutator of current operators.
More specifically, they can be expressed in terms of QPC-free correlation functions [$c^\mathcal{R}_j$, $c^\mathcal{A}_j$ and $c^\mathcal{K}_j$ of Eq.~\eqref{eq:free_c_functions_space}],
\begin{equation}
\begin{aligned}
    S^A_{ud} (x,y,0) & = \frac{1}{4\pi^2} c^\mathcal{R}_u (x,0,0) c^\mathcal{R}_d (y,0,0) S_\mathcal{T} (0) \\
    & = - S_\mathcal{T} (0) ,\\
    S^C_{ud} (x,y,0) & = \lim_{\omega \to 0} \frac{1}{\pi} \left[ c^\mathcal{K}_u (x,0,\omega) c^\mathcal{R}_d (y,0,-\omega) f_C (-\omega) \right.\\
    &\left.  - c^\mathcal{R}_u (x,0,\omega) c^\mathcal{K}_d (y,0,-\omega) f_C (\omega) \right]\\
    & =  i\frac{2 k_B }{\hbar} (T_u + T_d) \int_0^\infty \!\!dt \ t\  \big\langle [\hat{I}_\mathcal{T} (t) , \hat{I}_\mathcal{T} (0)] \big\rangle,\\
    f_C (\omega) & = \frac{1}{4\pi}\int_0^\infty dt \left( e^{i\omega t} - 1 \right) \big\langle [\hat{I}_\mathcal{T} (t) , \hat{I}_\mathcal{T} (0)] \big\rangle.
\end{aligned}
\label{eq:cross_correlations}
\end{equation}
where the communication between two involved edges is included by the noise $S_\mathcal{T}$, and the function $f_C$. The latter is proportional to the correlation of the tunneling current operator, with $\left[, \right]$ referring to the commutation operation.
In Eq.~\eqref{eq:cross_correlations}, correlations are evaluated with respect to the full Hamiltonian $H$, thus being strictly valid to all orders of the tunneling amplitude $\xi$, beyond perturbative regimes.

Now, following Eqs.~\eqref{eq:noise_relation} and \eqref{eq:cross_correlations}, we arrive at an important intermediate result:
\begin{equation}
\begin{aligned}
& S_{ud} (0) = \delta S_{ud} (0) + s_{ud} = - S_\mathcal{T} (0) \\
& + i \frac{2 k_B (T_u + T_d)}{\hbar} \int_0^\infty dt \ t\  \big\langle [\hat{I}_\mathcal{T} (t) , \hat{I}_\mathcal{T} (0)] \big\rangle,
\end{aligned}
\label{eq:srl}
\end{equation}
which provides us with the relation between two types of fluctuations, i.e., the cross correlation $S_{ud}$ and the tunneling current noise $S_\mathcal{T}$.
However, the physical significance of the integral of the second line of Eq.~\eqref{eq:srl} remains currently unclear, though this term is clearly connected to the conductance through the tunneling QPC (see, e.g., Refs.~\cite{fractional_statistics_theory_2016} for the leading-order tunneling current in the weak tunneling limit).
A more explicit relation between these two lines and the conductance will be provided shortly in Sec.~\ref{sec:fc_derivations}.

Now we move to consider autocorrelations, which equals $S_{\alpha \alpha} = S_{\alpha \alpha}^A + S_{\alpha \alpha}^C + s_{\alpha \alpha}$, where
\begin{equation}
\begin{aligned}
    S_{\alpha \alpha}^A & = - \lim_{\omega \to 0}\frac{\omega^2}{\pi} c^\mathcal{R}_\alpha (x,0,\omega) c^\mathcal{R}_\alpha (x,0,-\omega) \frac{S_\mathcal{T} (\omega) }{4\pi}\\
    & = S_\mathcal{T}(0),\\
    S_{\alpha \alpha}^C & =  \lim_{\omega \to 0}-\frac{\omega^2}{\pi} \left[ c^\mathcal{K}_\alpha (x,0,\omega) c^\mathcal{R}_\alpha (x,0,-\omega) f_C (-\omega) \right.\\
    & \left.  + c^\mathcal{R}_\alpha (x,0,\omega) c^\mathcal{K}_\alpha (x,0,-\omega) f_C (\omega) \right]\\
    & = - i\frac{4 k_B T_\alpha}{\hbar} \int_0^\infty dt \ t\  \big\langle [\hat{I}_\mathcal{T} (t) , \hat{I}_\mathcal{T} (0)] \big\rangle ,
\end{aligned}
\label{eq:auto_noise_ac}
\end{equation}
for $\alpha =$ $u$ and $d$.
We have thus obtained the immediate relation between autocorrelation and the tunneling current noise [akin to Eq.~\eqref{eq:srl}], as
\begin{equation}
\begin{aligned}
    &S_{\alpha \alpha} (x,x,0)  = S_\mathcal{T} + 4 \frac{\nu e^2}{h} k_B T_\alpha \\
    & -4 i \frac{k_B T_\alpha}{\hbar} \int_0^\infty dt\  t\  \big\langle [\hat{I}_\mathcal{T} (t) , \hat{I}_\mathcal{T} (0)] \big\rangle.
\end{aligned}
\label{eq:sauto}
\end{equation}
Eq.~\eqref{eq:sauto} depends explicitly only on the temperature $T_\alpha$ of the corresponding channel $\alpha$.
Actually, its dependence on the temperature of the other channel is implicitly contained by the correlation between the tunneling current operator.

Both Eqs.~\eqref{eq:srl} and \eqref{eq:sauto} contain the correlation that involves the commutator between tunneling current operators at different moments.
It is thus necessary to relate this correlation function to observables, to finally arrive at the non-equilibrium fluctuation relation given by Eq.~\eqref{eq:noise_conductance_relation}.

Before moving on, we stress that the strictly derivations of this section are obtained in ideal systems with the tunneling Hamiltonian Eq.~\eqref{eq:ht}.
Nevertheless, these results, especially Eqs.~\eqref{eq:cross_correlations} and \eqref{eq:auto_noise_ac}, can be extended to practical systems where tunnelings either occur in a finite-size area or are influenced by interaction factors; cf. Appendix~\ref{app:finite_size_dyson} for details.

\subsection{Strict relation between tunneling current noise and correlation functions}
\label{sec:fc_derivations}

The second line of both Eqs.~\eqref{eq:srl} and \eqref{eq:sauto} involve correlation of the commutator of tunneling operators.
Considering the existence of two edges, the small-frequency fluctuation-dissipation theorem Eq.~\eqref{eq:sg_relation}, after considering Eq.~\eqref{eq:noise_difference}, becomes simplified into,
\begin{equation}
\begin{aligned}
 &\lim_{\omega \to 0} \frac{\mathcal{C}_{ud} (\omega) - \mathcal{C}_{du} (-\omega)}{\omega}\\
    =& -\lim_{\omega \to 0} \int dt \frac{e^{i\omega t}}{\omega} \Big\langle \left[ \hat{I}_\mathcal{T} (0), \hat{I}_\mathcal{T} (t) \right] \Big\rangle\\
    = & 2i  \int_0^\infty dt \ t\  \Big\langle \left[ \hat{I}_\mathcal{T} (t), \hat{I}_\mathcal{T} (0) \right] \Big\rangle.
\end{aligned}
\label{eq:st_cross}
\end{equation}
With Eq.~\eqref{eq:st_cross}, the integral in the last two lines of Eqs.~\eqref{eq:srl} and \eqref{eq:sauto} can be alternatively written as [following also Eq.~\eqref{eq:dc_conductance}]
\begin{equation}
\begin{aligned}
    &2i \int_{0}^\infty dt \ t\  \langle [I_\mathcal{T} (t) , I_\mathcal{T} (0)] \rangle
    = \lim_{\omega\to 0} \frac{\mathcal{C}_{ud} (\omega) - \mathcal{C}_{du} (-\omega)}{\omega}\\
    & = \hbar \lim_{\omega\to 0} \left[ \mathcal{G}_{ud} (\omega) + \mathcal{G}_{du} (-\omega)\right]
    =\hbar \frac{\partial I_\mathcal{T}}{\partial V}  ,
\end{aligned}
\label{eq:fc_equality}
\end{equation}
where the last equality of the second line takes the extended fluctuation-dissipation theorem, Eq.~\eqref{eq:sg_relation}.
The last line on the other hand is the direct consequence of its definition, i.e., the first line of Eq.~\eqref{eq:response_definition}.

With Eq.~\eqref{eq:fc_equality}, and two intermediate relations of Eqs.~\eqref{eq:srl} and \eqref{eq:sauto}, we finally arrive at the relation between finite-frequency noises, and the matrix elements of tunneling-current conductance, i.e.,
\begin{equation}
\begin{aligned}
S_{ud} (0) & = - S_\mathcal{T} (0) \\
& + i \frac{2 k_B (T_u + T_d)}{\hbar} \!\int_0^\infty\!\! dt\ t\  \big\langle [\hat{I}_\mathcal{T} (t) , \hat{I}_\mathcal{T} (0)] \big\rangle,\\
& = - S_\mathcal{T} (0) + k_B (T_u + T_d) \frac{\partial I_\mathcal{T}}{\partial V },
\end{aligned}
\label{eq:cross_correlation_noise}
\end{equation}
for the cross correlation, and
\begin{equation}
\begin{aligned}
    S_{\alpha \alpha} (x,x,0) & = S_\mathcal{T} (0) + 4 \frac{\nu e^2}{h} k_B T_\alpha \\
    & -4 i \frac{k_B T_\alpha}{\hbar} \int_0^\infty dt\  t\  \big\langle [\hat{I}_\mathcal{T} (t) , \hat{I}_\mathcal{T} (0)] \big\rangle\\
    & = S_\mathcal{T}(0) + 4 G_0 k_B T_\alpha -2 k_B T_\alpha \frac{\partial I_\mathcal{T}}{\partial V },
\end{aligned}
\label{eq:auto_correlation_noise}
\end{equation}
for the autocorrelations. Eqs.~\eqref{eq:cross_correlation_noise} and \eqref{eq:auto_correlation_noise} perfectly agree with our major result, i.e., that given by Eq.~\eqref{eq:noise_conductance_relation}.

For the convenience of the readers, within this section, the evaluation of the zero-frequency expressions involves the approximations below, i.e.,
\begin{equation}
\begin{aligned}
    & \lim_{\omega = 0} \!\left[ \mathcal{G}_{ud} (\omega) + \mathcal{G}_{du} (-\omega)\right] = \frac{\partial I_\mathcal{T}}{\partial V } 
    ,\\
    & \lim_{\omega = 0} \coth\!\left( \!\frac{\hbar \omega}{2 k_B T} \!\right) \!\left[ \mathcal{G}_{ud} (\omega) \!+\! \mathcal{G}_{du} (-\omega)\right] \!=\!  \frac{2 k_B T }{\hbar \omega} \frac{\partial I_\mathcal{T}}{\partial V }
    ,\\
    & \lim_{\omega = 0} \nu \frac{e^2 w}{\pi} \coth\left( \frac{\hbar \omega}{2 k_B T_\alpha} \right) = 4\frac{\nu e^2}{h} k_B T_\alpha = 4 G_0 k_B T_\alpha,
\end{aligned}
    \label{eq:f_c-term-final}
\end{equation}
where the first line is obtained following the definition of the current response function, Eq.~\eqref{eq:response_definition}.
The second line, reflecting the autocorrelation for independent channels, is instead the Johnson–Nyquist noise.

Before ending this section, we stress that Eqs.~\eqref{eq:cross_correlations} and \eqref{eq:auto_noise_ac}, as the starting point of this subsection, are only valid for an ideal QPC.
For more practical situations, modifications are required on both equations, as the tunneling current operator should instead include tunneling events that occur within the entire transmission area.
Nevertheless, Eq.~\eqref{eq:st_cross} remains valid, after replacing the tunneling current operator, $\hat{I}_\mathcal{T}$ of an ideal case [i.e., Eq.~\eqref{eq:current_op}] by the more general expressions, i.e., $\hat{I}_\mathcal{T}^\text{eff}$ of Eq.~\eqref{eq:it_eff}.
Conclusion above requires the size of the transmission area to be much smaller than the thermal equilibration length, but much larger than the thermal diffusion length --- under which the temperatures and biases of tunneling-participating particles are well defined.
More details are provided in Appendix~\ref{app:finite_size_dyson}.

\section{Influence from a voltage-dependent transmission amplitude}
\label{sec:voltage_dependence}

Importantly, derivations of Sec.~\ref{sec:noise_conductance_relation} do not pre-assume the expression of the operator $\hat{I}_\mathcal{T}$, such that they are generically applicable to a wide range of system setups.
From this perspective, both Eq.~\eqref{eq:sg_relation}, obtained from generalized linear response, and our major achievements, Eq.~\eqref{eq:noise_conductance_relation}, are valid even if the transmission amplitude is momentum dependent [cf. Eqs.~\eqref{eq:ht_eff} and Eq.~\eqref{eq:ht_extended} of the Appendix~\ref{app:momentum_dependent}].
Actually, during these derivations, it is only assumed that $\xi$ has no dependence on the values of voltages $V_u$ and $V_d$: otherwise another contribution should be introduced into Eq.~\eqref{eq:interaction_eom}.
This assumption, while being experimentally much more realistic than a constant $\xi$, is, however, not always true, following observations of e.g., Ref.~\cite{kyrylo_FQHE_2024}.

In this section, we explicitly show the influence of the bias-dependency of the tunneling amplitude $\xi$ on our main result, Eq.~\eqref{eq:noise_conductance_relation_brief}.
We begin, before showing details, with the conclusion that the bias-dependency of $\xi$ only modifies Eq.~\eqref{eq:noise_conductance_relation_brief}, when the energy is not conserved at the tunneling area (notice that energy is conserved everywhere, except at the reservoir-edge contacts, or the tunneling area), due to e.g., electron-phonon coupling.
Other than that, Eq.~\eqref{eq:noise_conductance_relation_brief} remains valid, even when $\xi$ is bias-dependent.

\subsection{The general influence of the bias-dependent transmission amplitude}

As the simplification, in this work we assume $\xi [V_u, V_d ]$, meaning that $\xi$ does not depend on time $t$, to exclude high-frequency contributions.
This assumption perfectly applies to our currently interested setup, with DC drives at both sources. 
Importantly, even when $\xi$ is bias dependent, Eq.~\eqref{eq:interaction_eom} remains valid.
However, the momentum-dependence of $\xi$ now leads to an extra contribution when performing the differentiation over the bias ($V_u$ or $V_d$).
More specifically, the Hamiltonian now contains two bias-dependent contributions, given by
\begin{equation}
\begin{aligned}
    \delta H
    & = \delta V_j \left\{\hat{Q}_j + \frac{\partial \xi [V_u, V_d]}{\partial V_j}  \frac{1}{\xi} H_\mathcal{T}\right\},
\end{aligned}
    \label{eq:deltah_bias_dependent}
\end{equation}
where the second term is the consequence of the voltage-dependence of the tunneling Hamiltonian $H_\mathcal{T}$.
In Eq.~\eqref{eq:deltah_bias_dependent}, we have used the fact that $\partial \hat{I}_\alpha/\partial V_{\alpha'} = 0 $, when $\alpha \neq \alpha'$ [as current operators, cf. Eq.~\eqref{eq:charge_current}, do not explicitly depend on the applied biases].
As stressed at the beginning of this section, in this work we focus on the DC limit, and for later convenience, we define the quantity here.
\begin{equation}
    r_\alpha (V_{u0}, V_{d0}) \equiv \frac{\partial \ln \xi [V_u(t), V_d (t)]}{\partial V_\alpha}\Big|_{V_u (t)=V_{u0}, V_d (t)=V_{d0}},
    \label{eq:rj_definition}
\end{equation}
As a result of functional differentiation in the DC limit.
When $ r_j =0$, the transmission amplitude $\xi$ is bias-independent, where the results reduce to those of the previous section.

With Eq.~\eqref{eq:deltah_bias_dependent}, equation \eqref{eq:response_definition} becomes modified into
\begin{equation}
\begin{aligned}
    & \mathcal{G}^\text{mod}_{\alpha\alpha'} (t,t')\equiv\frac{\partial \mathcal{I}_\alpha (t)}{\partial V_{\alpha'} (t')}\Bigg|_{\frac{
    \partial \xi}{\partial V_{\alpha'}}  \neq 0} \\
    & = -\frac{i}{\hbar} \Theta (t-t') \Big\langle \left[ \hat{\mathcal{I}}_\alpha (t), \hat{Q}_{\alpha'} (t') + r_{\alpha'} H_\mathcal{T}(t') \right] \Big\rangle,
\end{aligned}
\label{eq:gv_kj}
\end{equation}
where we have replaced the differentiation over $\xi$ [cf. Eq.~\eqref{eq:rj_definition}] with $r_\alpha$ in the DC limit. In Eq.~\eqref{eq:gv_kj} we highlight the dependence of $\xi$ on the bias by the superscript ``mod''.
For later convenience, we decompose the generalized conductance matrix elements into two contributions, $\mathcal{G}^\text{mod}_{\alpha\alpha'} = \mathcal{G}_{\alpha\alpha'} + \mathcal{G}^\text{extra}_{\alpha\alpha'}$, with $\mathcal{G}_{\alpha\alpha'}$ being the contribution when $\xi$ is bias-independent, and
\begin{equation}
\begin{aligned}
    \mathcal{G}^\text{extra}_{\alpha\alpha'}(t,t') = - r_{\alpha'} \frac{i}{\hbar} \Theta (t-t') \Big\langle \left[ \hat{\mathcal{I}}_\alpha (t),  H_\mathcal{T}(t') \right] \Big\rangle,
\end{aligned}
\label{eq:gextra}
\end{equation}
as the extra contribution induced by the voltage-bias dependence of $\xi$.
To proceed, we differentiate $\mathcal{G}^\text{extra}_{\alpha\alpha'}(t,t')$ over $t'$, leading to
\begin{equation}
\begin{aligned}
    & \hbar \frac{\partial \mathcal{G}^\text{extra}_{\alpha\alpha'}(t,t')}{\partial t'}  = i\delta (t-t') r_{\alpha'} \Big\langle \left[ \hat{\mathcal{I}}_\alpha (t),  H_\mathcal{T}(t') \right] \Big\rangle\\
    & - i \Theta (t-t') r_{\alpha'} \Bigg\langle \left[ \hat{\mathcal{I}}_\alpha (t),  \frac{\partial H_\mathcal{T}(t')}{\partial t'} \right] \Bigg\rangle.
\end{aligned}
\label{eq:g_extra}
\end{equation}
Notice that in Eq.~\eqref{eq:g_extra}, contribution of the first line vanishes after taking $t = t'$.
Indeed, with the same argument in time, the current operator $\hat{I}_\alpha$ at the downstream of the QPC commutes with the tunneling Hamiltonian $H_\mathcal{T}$, $[\hat{I}_\alpha (x> 0, t),  H_\mathcal{T}(t) ] = 0$, as the tunneling occurs only at the QPC.
The second line of Eq.~\eqref{eq:g_extra}, on the other hand, can be evaluated following the equation of motion relation,
\begin{equation}
\begin{aligned}
     \frac{\partial H_\mathcal{T}(t')}{\partial t'}  & = \frac{i}{\hbar} \left[ H, H_\mathcal{T} \right] \\
     & = \frac{i}{\hbar} \left[ H_0 + V_u \hat{Q}_u + V_d \hat{Q}_d, H_\mathcal{T}  \right]\\
     & = \frac{i}{\hbar} \left[ H_0 + V_u \hat{Q}_u + V_d \hat{Q}_d, H  \right]\\
     & = - \hat{\mathcal{J}}_u - \hat{\mathcal{J}}_d,
\end{aligned}
\label{eq:ht_differentiation}
\end{equation}
where $\hat{\mathcal{J}}_\alpha \equiv i[H, H_\alpha + V_\alpha \hat{Q}_\alpha]/\hbar$ refers to the energy current that flows \textit{into} the edge $\alpha$.
Here the bias is included into the definition of $\hat{\mathcal{J}}_\alpha$, such that the energy of the ``Laughlin surface'' when $V_u = V_d = 0$ is chosen as the reference of energy.
Notice that the expectation of Eq.~\eqref{eq:ht_differentiation} should vanish, when the tunneling area does not allow the absorption or distribution of energies. This is different from the charge current operators, where $\hat{I}_u + \hat{I}_d =0$ due to the requested charge conservation through out the system of this work.

With Eqs.~\eqref{eq:gv_kj} and \eqref{eq:ht_differentiation}, we arrive at the conductance matrix elements in the time domain,
\begin{equation}
\begin{aligned}
    & \hbar \left[\frac{\partial \mathcal{G}^\text{mod}_{\alpha\alpha'} (t,t')}{\partial t'} \!-\! \frac{\partial \mathcal{G}^\text{mod}_{\alpha' \alpha} (t',t)}{\partial t}\right] = i \mathcal{C}_{\alpha\alpha'}(t,t') \!-\! i \mathcal{C}_{\alpha' \alpha} (t',t) \\
    & \quad\quad\quad-i \left\{\Theta (t-t') r_{\alpha'} \left[\mathcal{M}_{\alpha u}(t,t') + \mathcal{M}_{\alpha d}(t,t') \right] \right. \\
    &\quad\quad\quad\left. - \Theta (t'-t) r_\alpha \left[ \mathcal{M}_{\alpha' u}(t',t) + \mathcal{M}_{\alpha' d}(t',t)  \right] \right\},
\end{aligned}
\label{eq:generalized_fluctuation_theorem}
\end{equation}
where
\begin{equation}
    \mathcal{M}_{\alpha' \alpha} (t,t') \equiv \big\langle  \hat{\mathcal{J}}_\alpha (t') \hat{\mathcal{I}}_{\alpha'} (t) \big\rangle - \big\langle \hat{\mathcal{I}}_{\alpha'} (t) \hat{\mathcal{J}}_\alpha (t')  \big\rangle,
\end{equation}
refers to the ``mixed noise''~\cite{MatteoSciPhys25} between tunneling charge and heat currents.

\subsection{The influence of the bias-dependent transmission of the two-edge structure}

For the two-edge structure (i.e., Fig.~\ref{fig:model}) under current consideration, $\hat{\mathcal{I}}_u = - \hat{\mathcal{I}}_d = \hat{\mathcal{I}}_\mathcal{T} $, and we can introduce another energy-current operator $\mathcal{J} \equiv \hat{\mathcal{J}}_u + \hat{\mathcal{J}}_d $, referring to the energy absorption within the tunneling area.
Eq.~\eqref{eq:generalized_fluctuation_theorem} then simplifies into
\begin{equation}
\begin{aligned}
    & \hbar \left[\frac{\partial \mathcal{G}^\text{mod}_{\alpha\alpha'} (t,t')}{\partial t'} \!-\! \frac{\partial \mathcal{G}^\text{mod}_{\alpha' \alpha} (t',t)}{\partial t}\right] = i \mathcal{C}_{\alpha\alpha'}(t,t') \!-\! i \mathcal{C}_{\alpha' \alpha} (t',t) \\
    & -i \left[\Theta (t-t') r_{\alpha'} \varepsilon_{\alpha} \mathcal{M}(t,t')  - \Theta (t'-t) r_\alpha \varepsilon_{\alpha'} \mathcal{M} (t',t) \right]\\
    & = -i r_\alpha \varepsilon_{\alpha'} \mathcal{M}(t,t') \text{sgn} (t-t')
\end{aligned}
\label{eq:generalized_fluctuation_theorem_two_edge}
\end{equation}
with $\varepsilon_u = -\varepsilon_d = -1$, $\text{sgn}(t-t')$ the sign function, and the newly defined ``mixed noise'',
\begin{equation}
\begin{aligned}
    \mathcal{M} & \equiv \big\langle  \hat{\mathcal{J}} (t') \hat{\mathcal{I}}_\mathcal{T} (t) \big\rangle - \big\langle \hat{\mathcal{I}}_\mathcal{T} (t) \hat{\mathcal{J}} (t')  \big\rangle.
\end{aligned}
\end{equation}
In the last line of Eq.~\eqref{eq:generalized_fluctuation_theorem_two_edge}, we have taken the simplification that $r_u = -r_d$, under the assumption that $\xi$ depends on the bias difference, $V_u - V_d$.

When the Hamiltonian is time-independent, Eq.~\eqref{eq:generalized_fluctuation_theorem_two_edge} in the frequency domain can be rewritten as
\begin{equation}
\begin{aligned}
    &\hbar \omega \left[ \mathcal{G}^{\text{mod}}_{\alpha\alpha'} (\omega ) +  \mathcal{G}^{\text{mod}}_{\alpha' \alpha} (-\omega ) \right] = \mathcal{C}_{\alpha \alpha'} (\omega ) - \mathcal{C}_{\alpha' \alpha} (-\omega) \\
    & - i r_\alpha \varepsilon_{\alpha'} \mathcal{M}(\omega),
\end{aligned}
\label{eq:gmod_final}
\end{equation}
where
\begin{equation}
\begin{aligned}
    \mathcal{M}(\omega) & \equiv \int_{-\infty}^\infty e^{-i\omega t} \mathcal{M}(t,0) \ \text{sgn} (t) ,
\end{aligned}
\end{equation}
is the Fourier transformation that involves both $\mathcal{M}(t,0)$ and the sign function.
The second line of Eq.~\eqref{eq:gmod_final}, arising from the bias-dependency of $\xi$, modifies Eq.~\eqref{eq:fc_equality}, concerning the tunneling-current correlation, into,
\begin{equation}
\begin{aligned}
    &2i \int_{0}^\infty dt \ t\  \langle [I_\mathcal{T} (t) , I_\mathcal{T} (0)] \rangle
    = \lim_{\omega\to 0} \frac{\mathcal{C}_{ud} (\omega) - \mathcal{C}_{du} (-\omega)}{\omega}\\
    & = \hbar \lim_{\omega\to 0} \left\{ \left[ \mathcal{G}^\text{mod}_{ud} (\omega) + \mathcal{G}^\text{mod}_{du} (-\omega)\right] + i r_u \frac{\mathcal{M} (\omega)}{\omega} \right\}
    \\
    & =\hbar \left[ \frac{\partial I_\mathcal{T}}{\partial V} + \chi(\xi,V_u,V_d) \right] ,
\end{aligned}
\label{eq:fc_bias_dep}
\end{equation}
in terms of an auxiliary function,
\begin{equation}
    \chi(\xi,V_u,V_d) \equiv i r_u  \lim_{\omega \to 0} \frac{\mathcal{M} (\omega)}{\omega}.
\end{equation}

We have now arrived at the conclusion that when the transmission amplitude $\xi$ is bias-dependent, the major result Eq.~\eqref{eq:noise_conductance_relation_brief} becomes modified into
\begin{equation}
\begin{aligned}
    S_\mathcal{T} (0) &= -S_\text{ud}(0) + k_B (T_u + T_d) \left[ G + \chi (\xi,V_u,V_d) \right],\\
    S_\mathcal{T}(0) &= S_{\alpha \alpha}(0) + 2 k_B T_\alpha \left[ G\! +\! \chi (\xi,V_u,V_d) \right] \!-\! 4 G_0 k_B T_\alpha.
\end{aligned}
\label{eq:modified_flucutation_relation}
\end{equation}

Before ending this section, we stress again that this extra correction in Eq.~\eqref{eq:modified_flucutation_relation} is only finite when the tunneling area allows the absorption or injection of energies through inelastic scatterings.
Upon energy conservation at the QPC, $\lim_{\omega \to 0}\mathcal{M} (\omega)/\omega \to 0$, and Eq.~\eqref{eq:noise_conductance_relation_brief} remains valid even when $\xi$ is bias-dependent.

\section{Discussions}
\label{sec:summary}

In this work, with advances introduced to Refs.~\cite{DolciniPRB05,BenaSafiPBR07,ines_proceedings_2009,ines_philippe}, targeting at investigating bosonized out-of-equilibrium backscattering models, we have obtained the non-equilibrium DC fluctuation relation Eq.~\eqref{eq:noise_conductance_relation_brief} for the tunneling between two in-equilibrium chiral Laughlin edges that are featured by different temperatures and voltage biases.
Based on our derivation, the obtained non-equilibrium fluctuation relation is strictly valid for all values of biases, temperatures, and transmission amplitudes through the QPC, under only two prerequisites: (i) The time-dependency of two applied voltages, $V_u$ and $V_d$, and (ii) Two chiral Laughlin edge states are independent from each other, except at (or near) the tunneling QPC.
Given these two prerequisites satisfied, our theory remains valid even when the transmission probability is momentum dependent [cf. Eq.~\eqref{eq:ht_eff}, and see Appendix~\ref{app:momentum_dependent} for more details], or bias-dependent (cf. Sec.~\ref{sec:voltage_dependence}), or if the system is under the influence of interaction.
Corresponding modifications are also introduced and discussed, 
for the more complicated case where the transmission is both bias-dependent and inelastic
(cf. Eq.~\eqref{eq:modified_flucutation_relation} of Sec.~\ref{sec:voltage_dependence}).

Our theory goes beyond previous related literatures~\cite{Kane:1994b,Fendley:1996,WangFeldmanPRL13,FeldmanMotyPRB17,OneHalfPRB24} from two major perspectives.
Firstly, in addition to Refs.~\cite{Kane:1994b,Fendley:1996,WangFeldmanPRL13,FeldmanMotyPRB17} that only study the relation between auto correlation and the tunneling current noise, our Eq.~\eqref{eq:noise_conductance_relation_brief} also establishes the relation between cross correlation and tunneling current noise, for Laughlin edges with all filling fractions $\nu$.
Secondly, in addition to Refs.~\cite{Kane:1994b,Fendley:1996,WangFeldmanPRL13,FeldmanMotyPRB17,OneHalfPRB24} where the transmission probability through the QPC is assumed as a constant, here we manage to show that the non-equilibrium fluctuation relations, Eq.~\eqref{eq:noise_conductance_relation_brief} is robust against various interruptions, including interactions, a momentum-dependent transmission amplitude [$\xi$ of Eq.~\eqref{eq:ht}], as well as a temperature difference between two involved edges.
The latter especially benefits experimental studies in the field of delta-$T$ noise.
In addition, for systems where the tunneling amplitude through the QPC depends on the bias,
Eq.~\eqref{eq:noise_conductance_relation_brief} remains valid as long as energy is conserved within the tunneling area.
Otherwise corrections are explicitly provided.
Considering the experimental relevance of these interrupting elements, our work greatly extended the range of applicability of the obtained non-equilibrium fluctuation relation, into practical experiments.

Before ending, we would like to mention that the present approach is complementary to non-equilibrium bosonization approaches~\cite{GutmanGefenMirlinPRL08,KovrizhinChalkerPRB09,LevkivskyiSukhorukovPRL09,GutmanGefenMirlinPRB10} to FQH edges, in particular to the recent formulation of Ref.~\cite{SpanslattParkMirlinX26}. Those approaches construct the Green functions and full counting statistics of edges prepared in non-thermal stationary states and then compute QPC tunneling currents and noises,typically perturbatively in the tunneling amplitude of the analyzed contact. Here, by contrast, we derive fluctuation relations between downstream current correlations and the local tunneling noise for voltage- and temperature-biased constrictions. The present identities do not compute the tunneling noise itself; rather, they determine how it is encoded in experimentally measured auto- and cross-correlations.

\textbf{\emph{Acknowledgments---}} The authors are grateful to Yuval Gefen and Kyrylo Snizhko for fruitful discussions.
GZ acknowledges the support from the National Natural Science Foundation of China under Grants No. 12674059, and the startup grant at Nanjing University.
IS acknowledges the support from the French National Research Agency (ANR) under Grant No. ANR-21-CE47-0012 (``QuSig4QuSense'').

\begin{appendix}

\section{Derivation of the tunneling current operator of Eq.~\eqref{eq:iud_general}}
\label{eq:iud_derivation}

In this section, we present the derivation of Eq.~\eqref{eq:iud_general}, which expresses the tunneling current in terms of the derivative of $H_\mathcal{T}$, without specifying its explicit form. Consequently, the results derived in this section are generally applicable and are not restricted to particular forms of the tunneling Hamiltonian.

Actually, the derivation of Eq.~\eqref{eq:iud_general} requires only two starting relations,
\begin{equation}
\begin{aligned}
    &\hat{Q}_\alpha = - \frac{\sqrt{\nu}}{2\pi} \int dx \ \partial_x \phi_\alpha (x),\\
    &\left[ \phi_\alpha (x) , \partial_{x'} \phi_\beta (x') \right] = 2\pi i \delta_{\alpha\beta} \delta (x-x'),
\end{aligned}
\label{eq:charge_integral}
\end{equation}
i.e., the charge operator in terms of the bosonic field operators, and their corresponding bosonic commutation relation.

To start with, the commutator of Eq.~\eqref{eq:charge_integral} directly leads to
\begin{equation}
\begin{aligned}
    \left[ \hat{Q}_\alpha, \phi_\beta (x) \right] & = -\frac{\sqrt{\nu}}{2\pi} \int dx' \ \left[ \partial_{x'} \phi_\alpha (x'), \phi_\beta (x) \right]\\
    & = i \sqrt{\nu} \delta_{\alpha\beta},
\end{aligned}
\end{equation}
indicating the the charge operator $\hat{Q}_\alpha$ and the bosonic operator $\phi_\alpha$ form into a canonical pair.
This feature leads further to,
\begin{equation}
    e^{i\lambda Q_\alpha } \phi_\beta e^{-i\lambda Q_\alpha } = \phi_\beta - \lambda \sqrt{\nu} \delta_{\alpha\beta},
    \label{eq:operator_shift}
\end{equation}
i.e., a shift in the bosonic operator, by the exponential of the corresponding charge operator, where $\lambda$ is a constant number.
This shift of the bosonic operator applies also to the tunneling Hamiltonian, leading to
\begin{equation}
\begin{aligned}
    e^{i\lambda Q_u } H_\mathcal{T} [\phi_u, \phi_d] e^{-i\lambda Q_u } & = H_\mathcal{T} [\phi_u - \lambda\sqrt{\nu}, \phi_d],\\
    [\hat{Q}_\alpha, H_\mathcal{T}] & = i \sqrt{\nu} \frac{\partial H_\mathcal{T}}{\partial \phi_\alpha} ,
\end{aligned}
\label{eq:charge_htun_comm}
\end{equation}
where we have taken, without the loss of generality, $\alpha = u$ in the first line, and the second line is obtained after differentiating the first line by $\lambda$, and taking $\lambda \to 0$ afterwards.
Notably, Eq.~\eqref{eq:charge_htun_comm} is obtained without specifying the explicit expression of $H_\mathcal{T}$.

With Eq.~\eqref{eq:charge_htun_comm}, we finally arrive at Eq.~\eqref{eq:iud_general} for the current operator, i.e.,
\begin{equation}
\begin{aligned}
    I_\mathcal{T} & = - \frac{\partial}{\partial t} \hat{Q}_u = i \left[ \hat{Q}_u, H_\mathcal{T} \right] = -\sqrt{\nu} \frac{\partial H_\mathcal{T}}{\partial \phi_u} \\
    & = \frac{\partial}{\partial t} \hat{Q}_d = i \left[ \hat{Q}_d, H_\mathcal{T} \right] = \sqrt{\nu} \frac{\partial H_\mathcal{T}}{\partial \phi_d} .
\end{aligned}
\end{equation}

\section{Basic concept of the generalized linear response theory}
\label{app:genearlized_linear_response}

In this section we introduce fundamental concepts of the framework of the generalized linear response theory.
Generally, we consider situations where the system Hamiltonian can be written as $H' = H_0 + \delta H$,
where $\delta H$ refers to the infinitesimal variation of the drive, that is introduced on top of the original Hamiltonian $H_0$ [for instance, but not limited to, Eq.~\eqref{eq:full_hamiltonian} of our model].
Unlike that of the linear response theory, here $H_0$ can refer to systems either in or out of equilibrium.
Within the framework of the generalized linear response, we take the \textit{interaction} picture, with $\delta H$ the ``interacting'' part.
Within this picture, evolution of wave functions is described by the equation of motion,
\begin{equation}
\begin{aligned}
    i\hbar \frac{\partial |\Psi_\text{int} (t) \rangle }{\partial t} & = U(t,-\infty) \delta H U^\dagger(t,-\infty) |\Psi_\text{int} (t) \rangle,
\end{aligned}
\label{eq:interaction_evolution}
\end{equation}
where $|\Psi_\text{int} (t) \rangle$ is the time-dependent wave function in the interaction picture, and
$U \equiv \exp(iH_0 t/\hbar)$ is the time-evolution operator defined with respect to $H_0$, the Hamiltonian Eq.~\eqref{eq:full_hamiltonian} before introducing the infinitesimal extra drive.
To the leading order of $\delta H$, the wave function becomes perturbatively
\begin{equation}
    |\Psi_\text{int} (t) \rangle \approx |\Psi_\text{int} (-\infty) \rangle - \frac{i}{\hbar} \int_{-\infty}^t dt' \delta H (t') |\Psi_\text{int} (-\infty) \rangle.
    \label{eq:equation_of_motion}
\end{equation}
Following Eq.~\eqref{eq:equation_of_motion}, 
for any operator $\hat{\mathcal{O}}$, its expectation value under the full Hamiltonian becomes approximately
\begin{equation}
\begin{aligned}
    & \langle \hat{\mathcal{O}} \rangle_{H'} (t)  \equiv \langle \Psi_\text{int} (t) | \hat{\mathcal{O}} (t) | \Psi_\text{int} (t) \rangle\\
    & \approx \langle \hat{\mathcal{O}} \rangle_{H_0} (t) \\
    & - \frac{i}{\hbar} \int_{-\infty}^t dt' \langle \Psi_\text{int} (-\infty) | \left[ \hat{\mathcal{O}} (t) ,\delta H (t')  \right]|\Psi_\text{int} (-\infty ) \rangle,
\end{aligned}
\label{eq:oh_expression}
\end{equation}
where $\langle \hat{\mathcal{O} }\rangle_{H'}$ and $\langle \hat{\mathcal{O}} \rangle_{H_0}$ refer to correlation functions evaluated with respect to $H'$ and $H_0$, respectively.
Again, in Eq.~\eqref{eq:oh_expression}, the time-dependent operator $\hat{\mathcal{O}} (t) \equiv \exp (iH_0 t/\hbar) \hat{\mathcal{O}} \exp (-iH_0 t/\hbar)$ is defined in the interaction picture, as all other operators and wave functions.

In the generalized linear response, $\delta H$ refers to an infinitesimal extra drive that can be written explicitly as $\delta H (t) = \sum_j \delta \hat{F}_j (t) \hat{N}_j$, where $\delta \hat{F}_j$ refers to the infinitesimal modification of the drive $\hat{F}_j$, that couples linearly to the corresponding canonical ``charge'' $\hat{N}_j$.
Within this linear assumption, we obtain the formula of the generalized linear response theory,
\begin{equation}
\begin{aligned}
    &\frac{\partial \hat{\langle O} \rangle (t)}{\partial \hat{F}_j(t')} = -\Theta (t-t')  \\
    & \times \Big\langle \Psi_\text{int} (-\infty) | \left[ O (t) ,\hat{N}_j (t')  \right]|\Psi_\text{int} (-\infty ) \Big\rangle,
    \end{aligned}
    \label{eq:linear_response_expansion}
\end{equation}
which describes the response of the operator expectation [i.e., $\langle O \rangle (t)$] to an infinitesimal modification of the drive, $\delta \hat{F}_j$.
Notice that Eq.~\eqref{eq:linear_response_expansion} contains a theta function in time, indicating the causality requirement, i.e., that the expectation value of $\hat{\mathcal{O}}$ at time $t$ only receives influences from moments before $t$.
In obtaining Eq.~\eqref{eq:linear_response_expansion}, we do not assume any detail of the full Hamiltonian $H$, except that $\delta H = \sum_j \delta \hat{F}_j \hat{N}_j$, i.e., a linear coupling between the generalized particle number $\hat{N}_j$ and the modification on the drive $\hat{F}_j$.
Actually, Eq.~\eqref{eq:linear_response_expansion} remains valid beyond the linear coupling regime, as long as the modification of the drive, $\langle\delta \hat{F}_j \rangle \ll \langle \hat{F}_j \rangle$, is a small quantity, and that the linear-order term remains finite.
This is exactly the case discussed by Eq.~\eqref{eq:deltah_bias_dependent} of Sec.~\ref{sec:voltage_dependence}, where the tunneling Hamiltonian $H_\mathcal{T}$ becomes dependent on the applied voltage bias.
Actually, in Eq.~\eqref{eq:deltah_bias_dependent}, the generalized particle number $\hat{N}_j$ contains all operators within the brackets.

\section{Momentum-dependent transmission amplitude}
\label{app:momentum_dependent}

In the main text, we mentioned that in practical systems, the tunneling amplitudes between two channels effectively depend on the momentum of involved anyonic operators.
In this section we explicitly present the basic idea of three mechanisms that can induce this dependence of tunneling amplitudes, on the momentum.
These mechanisms include: (i) an extended quantum point contact, (ii) the presence of a dot within the tunneling area, and (iii) the presence of interaction, either electron-phonon or electron-electron, near the QPC.
Before moving to specific examples, Hamiltonian of all these mechanisms can be incorporated by the effective Hamiltonian
\begin{equation}
\begin{aligned}
    H_\mathcal{T}^\text{eff} & = \iint dx dx' \xi^\text{eff} (x,x') \psi^\dagger_u (x) \psi_d(x') + h.c.\\
    & = \iint dx dx' h^\text{eff}_\mathcal{T} + h.c.,\\
    h^\text{eff}_\mathcal{T} & = \xi^\text{eff} (x,x') \psi^\dagger_u (x) \psi_d(x') + h.c.,
\end{aligned}
\label{eq:effective_tunneling_hamiltonian}
\end{equation}
with $\psi^\dagger_\alpha$ the minimum quasiparticle excitation of edge $\alpha$, $\xi^\text{eff} (x, x')$ a position-dependent transmission amplitude (from position $x$ to $x')$, and $h^\text{eff}_\mathcal{T}$ the effective local tunneling Hamiltonian ``density''.
Importantly, Eq.~\eqref{eq:effective_tunneling_hamiltonian} applies not only to systems containing a finite-size tunneling area, but also to those where tunneling is influenced by interactions, thus becoming momentum dependent.
Crucially, in all these setups, the extended tunneling area is assumed to be much smaller than the size of the equilibration length, since otherwise particles participating the tunneling should have position-dependent temperatures.

Below we illustrate three specific cases, from Sec.~\ref{sec:extended_qpc} to Sec.~\ref{sec:interacting_qpc}, as examples where Eq.~\eqref{eq:effective_tunneling_hamiltonian} is applicable.

\subsection{An extended quantum point contact}
\label{sec:extended_qpc}

Here, by an ``extended'' point contact, we refer to that where tunneling between two channels can occur between points within a finite-size area [cf. Fig.~\ref{fig:momentum_dependent_transmission}(a)].
Most generically, tunneling between the QPC can be described by the Hamiltonian
\begin{equation}
\begin{aligned}
    & H_\mathcal{T}^\text{ext}  = \iint dk dk' \xi^\text{ext}_{k,k'} c_{u,k}^\dagger c_{d,k'} + h.c.,\\
    & = \frac{1}{2\pi} \iint dk dk' \iint dx dx' \xi^\text{ext}_{k,k'} e^{i(kx - k'x')}\psi_u^\dagger (x) \psi_d (x') \\
    & = \iint dx dx' \xi^\text{ext} (x,x') \psi^\dagger_u (x) \psi_d (x') + h.c.,
\end{aligned}
\label{eq:ht_extended}
\end{equation}
with the superscript ``ext'' referring to the extended tunneling QPC, and $c^\dagger_{\alpha, k}$ 
refers to the anyonic operator with momentum $k$, in channel $\alpha$,
\begin{equation}
\begin{aligned}
    c^\dagger_{\alpha, k} & \equiv \frac{1}{\sqrt{2\pi}} \int dx e^{ikx} \psi^\dagger_\alpha (x),\\
    c_{\alpha, k} & \equiv \frac{1}{\sqrt{2\pi}} \int dx e^{-ikx} \psi_\alpha (x),
\end{aligned}
    \label{eq:ck}
\end{equation}
defined as the Fourier transform of quasiparticle operators in real space.
Following the last line of Eq.~\eqref{eq:ht_extended}, the tunneling amplitude ``density'', $\xi^\text{ext} (x,x')$, equals
\begin{equation}
    \xi^\text{ext} (x,x') = \frac{1}{2\pi} \iint dk dk' \xi^\text{ext}_{k,k'} e^{i(kx - k'x')},
    \label{eq:zeta_xx}
\end{equation}
or inversely,
\begin{equation}
    \xi^\text{ext}_{k,k'} = \frac{1}{2\pi} \iint dx dx' \xi^\text{ext} (x,x') e^{-i(kx - k'x')}.
    \label{eq:zeta_kk}
\end{equation}
For the simplest situation, where $\xi^\text{ext}_{k,k'} = \xi /2\pi$, its expression in real space becomes $\xi^\text{ext}(x-x') \to \xi \delta (x) \delta (x')$, where the extended Hamiltonian Eq.~\eqref{eq:ht_extended} reduces to the ideal one, Eq.~\eqref{eq:ht}, after bosonization.
In general, however, for a quantum point contact with an extended area, the tunneling amplitude $\xi^\text{ext} (x, x')$ is non-zero in a finite-size area.
For this case, following Eq.~\eqref{eq:zeta_kk}, $\xi^\text{ext}_{k,k'}$ becomes effectively momentum dependent.

\subsection{Quantum-dot influenced tunneling}

\begin{figure}
    \centering
    \includegraphics[width=0.95\linewidth]{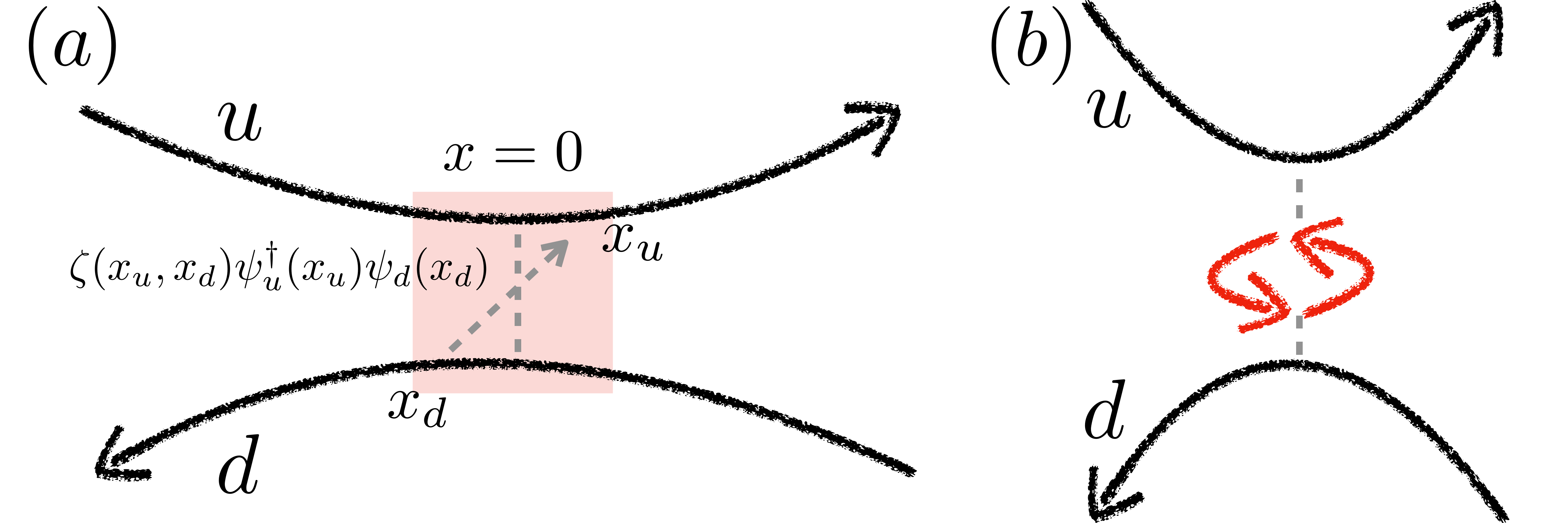}
    \caption{Two potential setups where tunneling between two channels, 1 and 2, effectively depends on the momentum of tunneling quasiparticles.
    (a) The setup where the tunneling between two channels can occur between positions within a finite-size area (the red area), not only at the $x = 0$ point.
    (b) The setup involving a quantum dot (the red arrows) between two chiral channels.
    Although initially tunneling occurs only at one point, $x = 0$, the tunneling amplitude becomes momentum-dependent, after integrating out the dot degree of freedom.
    }
    \label{fig:momentum_dependent_transmission}
\end{figure}

Alternatively, we can consider the situation where the tunneling between two chiral channels is influenced by a quantum dot, cf. Fig.~\ref{fig:momentum_dependent_transmission}(b).
For this case, the tunneling between two chiral channels and the anti-dot can be modeled by the Hamiltonian
\begin{equation}
\begin{aligned}
    H_\mathcal{T}^\text{dot} & = \xi_u \psi_u^\dagger (0) d + \xi_d \psi_d^\dagger (0) d + h.c.,
\end{aligned}
\label{eq:ht_dot}
\end{equation}
where $d$ and $d^\dagger$ refer to the annihilation and creation operators, respectively, of the only tunneling-relevant dot state.
These operators, when acting on the corresponding dot states $|0\rangle$ and $|1\rangle$, lead to
\begin{equation}
\begin{aligned}
    & d |0\rangle = d^\dagger |1 \rangle = 0,\\
    & d |1 \rangle = |0\rangle, \ d^\dagger |0\rangle = |1 \rangle.
\end{aligned}
\end{equation}
We further assume that the dot state is away from resonance, such that (without loss of generality) only the state $|0\rangle$ is allowed as the real state.

For this case, one can obtain the effective tunneling between channels $u$ and $d$, after removing the dot degree of freedom with the Schrieffer-Wolff transformation.
To begin with, we rewrite Eq.~\eqref{eq:ht_dot} in terms of operators in momentum representation [cf. Eq.~\eqref{eq:ck}], leading to
\begin{equation}
\begin{aligned}
    H_\mathcal{T}^\text{dot} & = \xi_u \frac{1}{\sqrt{2\pi}} \int dk e^{-ik x} c^\dagger_{u,k} d + h.c.,\\
    & + \xi_d \frac{1}{\sqrt{2\pi}} \int dk e^{-ik x} c^\dagger_{d,k} d + h.c..
\end{aligned}
\label{eq:ht_dot_k}
\end{equation}
Assuming that the anyonic state with momentum $k$ has the energy $\epsilon_k$, then after the Schrieffer-Wolff transformation, the tunneling between channels $u$ and $d$ can be described by the effective tunneling Hamiltonian
\begin{equation}
\begin{aligned}
    H_\mathcal{T}^\text{dot,eff} = \frac{1}{2\pi}\iint dk dk' \xi_{k,k'}^\text{eff} c_{u,k}^\dagger c_{d,k'} + h.c.,
\end{aligned}
\end{equation}
which share the structure of the first line of Eq.~\eqref{eq:ht_extended}.
Here, the effective tunneling amplitude $\xi_{k,k'}^\text{eff}$ equals
\begin{equation}
    \xi_{k,k'}^\text{eff} = \frac{ \xi_u \xi_d}{\epsilon_{k'} - \epsilon_d },
\end{equation}
where the denominator equals the difference in energy, between that of the initial ($\epsilon_{k'}$, with the dot empty), and that of the middle ($\epsilon_d$, with the dot occupied) states.

As the consequence, in this dot-influenced tunneling model, although in the original Hamiltonian anyons in both channels only occur at a single point, the effective Hamiltonian, after including the anti-dot into consideration, becomes position-dependent, with the effective coupling parameter in space:
\begin{equation}
    \xi^\text{eff} (x,x') = \frac{1}{2\pi} \iint dk dk' \xi_{k,k'}^\text{eff} e^{i(kx - k'x')}.
    \label{eq:zeta_eff}
\end{equation}
We stress again that Eq.~\eqref{eq:zeta_eff} is only an effective Hamiltonian obtained after taking into consideration the function of the dot.
Strictly speaking, however, the tunneling occurs only at a single point of each channel.

\subsection{Interaction at the QPC}
\label{sec:interacting_qpc}

As the third example, the tunneling amplitude $\xi_{k,k'}^\text{eff}$ can be induced by interactions, either electron-phonon or electron-electron, at the tunneling QPC.

As the starting point, following the tunneling current operator Eq.~\eqref{eq:current_op}, the correlation between anyonic operators, e.g., $\langle \exp [-i\sqrt{\nu}\phi_u (t)] \exp [i\sqrt{\nu}\phi_u (0)] \rangle$, is critical in the evaluation of quantum transport, including the charge and heat tunneling currents and the associated noises.
For an ideal QPC (i.e., when tunneling is free from interactions, and occurs at a single-point), after including the prefactor $|\xi|^2$, the correlation function has the long-time feature
\begin{equation}
\begin{aligned}
    & |\xi|^2 \big\langle e^{-i\sqrt{\nu}\phi_u (t)} e^{i\sqrt{\nu}\phi_u (0)} \big\rangle \big\langle e^{i\sqrt{\nu}\phi_d (t)} e^{-i\sqrt{\nu}\phi_d (0)} \big\rangle \\
    = & |\xi|^2 \frac{l_c^{2\nu}}{( l_c + i t )^{2\nu}} \sim \frac{1}{t^{2\nu}},
\end{aligned}
    \label{eq:idea_scaling}
\end{equation}
where the scaling factor $2\nu$ equals twice the filling fraction of the corresponding fractional quantum Hall system.
This scaling factor, after the Fourier transformation, leads to the transmission probability at energy $\epsilon$, as
\begin{equation}
    D(k) = D_0 \left( \frac{\epsilon_{k}}{\epsilon_{k_0}} \right)^{2\nu - 1} \propto |\xi|^2 \epsilon_k^{2\nu - 1},
    \label{eq:d_epsilon}
\end{equation}
where $\epsilon_k$ equals the energy of quasiparticle excitation with momentum $k$, and $D_0$ refers to the transmission probability obtained at the initial energy $\epsilon_{k_0}$, which is determined by the system drive or the thermal fluctuation.

Practically, however, influenced by e.g., electron-electron or electron-phonon interactions, the scaling factor of Eq.~\eqref{eq:idea_scaling} does not necessarily equal the filling fraction. For instance, when anyonic tunnelings are coupled to an Ohmic bosonic bath (with Ohmic resistance $R$), the tunneling Hamiltonian becomes effectively~\cite{Ingold1992SUS}
\begin{equation}
    H_\mathcal{T}^\text{e-ph} = \frac{ \xi}{ ( 2 \pi l_c )^\nu } e^{i \sqrt{\nu} \phi_u } e^{ -i \sqrt{\nu} \phi_u } e^{i\varphi} + h.c.,
\end{equation}
where $\varphi$ has the dissipation-induced long-time dynamical feature
\begin{equation}
    \big\langle e^{i\varphi (t)} e^{-i\varphi (0)} \big\rangle \propto \frac{1}{t^{2r}},
    \label{eq:dissipative_correlation}
\end{equation}
where $r \equiv R/R_Q$ is the dimensionless strength of dissipation. Here $R_Q = h/e^2$ is the dissipation quanta.
Due to Eq.~\eqref{eq:dissipative_correlation}, in a dissipative system, correlation function Eq.~\eqref{eq:idea_scaling} has a different long-time feature, i.e.,
\begin{equation}
\begin{aligned}
    \big\langle e^{-i\sqrt{\nu}\phi_u (t)} e^{i\sqrt{\nu}\phi_u (0)} \big\rangle \big\langle e^{i\sqrt{\nu}\phi_d (t)} e^{-i\sqrt{\nu}\phi_d (0)} \big\rangle \big\langle e^{i\varphi (t)} e^{-i\varphi (0)} \big\rangle \\
    \propto \frac{1}{t^{2\nu + 2r}},
\end{aligned}
    \label{eq:scaling_dissipative}
\end{equation}
where the scaling power factor changes from $2\nu$ to $2\nu + 2 r$, due to the coupling of anyonic tunneling to an outer bosonic environment.
Similarly, the interaction-induced modification on the scaling power factor is reported by Ref.~\cite{RosenowHalperinPRL02}, where an electron-phonon coupling is introduced along the entire edge.

After the Fourier transformation, Eq.~\eqref{eq:scaling_dissipative} leads to another momentum-dependence of the transmission probability, i.e.,
\begin{equation}
    D^\text{e-ph} (k) \propto |\xi|^2 \epsilon_k^{2\nu' - 1}
    \label{eq:d_e-ph}
\end{equation}
where $\nu' \equiv \nu + r$ receives the influence from the dissipative environment, and the superscript ``e-ph'' highlights the influence from the electron-phonon interaction.

Eqs.~\eqref{eq:scaling_dissipative} and \eqref{eq:d_e-ph} importantly disclose that in practical experiments, the scaling factor can deviate from the ideal (i.e., interaction free) theoretical prediction after taking interactions into consideration: a fact that can explain the deviation between theoretical expectations and practical experimental observations.
Alternatively, the interaction-induced anomalous scaling factor can be absorbed by redefining the constant tunneling amplitude $\xi$ into an effective function, $\xi^\text{e-ph} (k)$, i.e.,
\begin{equation}
    D^\text{e-ph} (k) \propto |\xi^\text{e-ph} (k)|^2 \epsilon_k^{2\nu - 1},
    \label{eq:xi_v_dependent}
\end{equation}
where
\begin{equation}
    \xi^\text{e-ph} (k) \propto \xi \epsilon_k^{2(\nu'-\nu)},
    \label{eq:momentum_dependent_xi}
\end{equation}
is the effective interaction-influenced tunneling amplitude.
Eq.~\eqref{eq:momentum_dependent_xi} discloses the fact that the influence of dissipative modes can be \textit{effectively} incorporated by introducing a voltage-dependent transmission amplitude, $\xi^\text{e-ph} (\epsilon)$.
This feature importantly also appears in systems where electrons Coulomb-interact at the QPC, cf. Ref.~\cite{PapaMacDonaldPRL04}.

\section{Derivation of intermediate expressions with generating functionals}
\label{app:intermediate_finite_size}

Eqs.~\eqref{eq:noise_relation}, \eqref{eq:cross_correlations}, and \eqref{eq:auto_noise_ac} are the essential starting equations to obtain our final result, i.e., Eq.~\eqref{eq:noise_conductance_relation_brief}, which establishes the relation between conductance and noise.
In this section, we derive these equations using the generating functional method~\cite{DolciniPRB05}.
More specifically, we consider two situations of the QPC, i.e., the ideal one and the more practical one, where tunneling through the QPC occurs at a single point and over a finite range of area (cf. Appendix~\ref{app:momentum_dependent}), in Appendices~\ref{app:ideal_collider} and \ref{app:finite_size_dyson}, respectively.
Derivations of the latter section, notably, involve current operators defined for an extended QPC.

\subsection{The ideal QPC}
\label{app:ideal_collider}

We begin with the ideal situation, where tunneling through the QPC can be described by the tunneling Hamiltonian Eq.~\eqref{eq:ht}, i.e., when the tunneling occurs at a single point and is free from any interaction.
We begin by introducing the QPC-free Green's functions,
\begin{equation}
\begin{aligned}
    g^\mathcal{A}_\alpha (x,t;x',t') & = -\theta (t'-t) \big\langle [ \phi_\alpha (x,t), \phi_\alpha (x',t') ] \big\rangle_0,\\
    g^\mathcal{R}_\alpha (x,t;x',t') & = \theta (t-t') \big\langle [ \phi_\alpha (x,t), \phi_\alpha (x',t') ] \big\rangle_0,\\
    g^\mathcal{K}_\alpha (x,t;x',t') & = \big\langle [ \phi_\alpha (x,t), \phi_\alpha (x',t') ] \big\rangle_0,
\end{aligned}
\label{eq:impurity_free_gfs}
\end{equation}
with their explicit expressions in the frequency domain,
\begin{equation}
\begin{aligned}
    g^\mathcal{A}_\alpha (x,x',\omega)
    & = - \frac{2\pi}{\omega} e^{- i \nu \omega (x - x')/v} \Theta  (x - x'),\\
    g^\mathcal{R}_\alpha (x,x',\omega) & = \frac{2\pi}{\omega} e^{- i \nu \omega (x - x')/v} \Theta [- (x - x')],\\
    g^\mathcal{K}_\alpha (x,x',\omega) & = \frac{2\pi}{\omega} e^{- i \nu \omega (x - x')/v} \coth \left( \frac{\hbar\omega}{2k_B T_\alpha} \right).
\end{aligned}
\label{eq:free_greens_functions_space}
\end{equation}
For the convenience of the derivations that follow, we also choose to define the Green's function matrix here,
\begin{equation}
g_\alpha = 
\begin{pmatrix}
g_\alpha^{++} & g_\alpha^{+-} \\
g_\alpha^{-+} & g_\alpha^{--} 
\end{pmatrix},
\label{eq:gf_matrix}
\end{equation}
in terms of four free-edge Green's functions.
These Green's functions are related to the retarded, advanced, and Keldysh Green's functions [cf. Eq.~\eqref{eq:free_greens_functions_space}], following
\begin{equation}
\begin{aligned}
    g^{++}_\alpha (x,t; x',t') & = \frac{1}{2} \left[ g^\mathcal{K}_\alpha (x,t; x',t') - g^\mathcal{R}_\alpha (x,t; x',t') \right.\\
    &\left. - g^\mathcal{A}_\alpha (x,t; x',t') \right],\\
    g^{+-}_\alpha (x,t; x',t') & = \frac{1}{2} \left[ g^\mathcal{K}_\alpha (x,t; x',t') + g^\mathcal{R}_\alpha (x,t; x',t') \right.\\
    & \left. - g^\mathcal{A}_\alpha (x,t; x',t') \right],\\
    g^{-+}_\alpha (x,t; x',t') & = \frac{1}{2} \left[ g^\mathcal{K}_\alpha (x,t; x',t') - g^\mathcal{R}_\alpha (x,t; x',t') \right.\\
    &\left. + g^\mathcal{A}_\alpha (x,t; x',t') \right],\\
    g^{--}_\alpha (x,t; x',t') & = \frac{1}{2} \left[ g^\mathcal{K}_\alpha (x,t; x',t') + g^\mathcal{R}_\alpha (x,t; x',t') \right.\\
    & \left. + g^\mathcal{A}_\alpha (x,t; x',t') \right].
\end{aligned}
\label{eq:gf_rotations}
\end{equation}
Clearly, only three out of four functions above are independently defined.
Notice that Eqs.~\eqref{eq:impurity_free_gfs} and \eqref{eq:free_greens_functions_space} basically share the structure of Eq.~\eqref{eq:free_c_functions_space} of the main text, except that the latter contains an extra differentiation over time, which removes the dependence of $\omega$ on the denominator in Eqs.~\eqref{eq:impurity_free_gfs}.

With current operators defined by Eq.~\eqref{eq:charge_current},
the cross correlation function, defined by Eq.~\eqref{eq:auto_cross_correlations} in terms of current operators, can be alternatively written as
\begin{widetext}
\begin{equation}
\begin{aligned}
    S_{ud} (\omega;x_u,x_d) \equiv &\frac{\nu e^2}{4\pi^2}\int_{-\infty}^\infty dt e^{i\omega t} \left[\big\langle \dot{\phi}_u (x_u, t) \dot{\phi}_d (x_d,0) \big\rangle  -  \big\langle \dot{\phi}_u (x_u, t) \rangle \langle \dot{\phi}_d (x_d,0) \big\rangle\right]\\
    + & \frac{\nu e^2}{4\pi^2}\int_{-\infty}^\infty dt e^{-i\omega t} \left[\big\langle \dot{\phi}_d (x_d,0) \dot{\phi}_u (x_u, t)  \big\rangle  -  \big\langle \dot{\phi}_u (x_u, t) \rangle \langle \dot{\phi}_d (x_d,0) \big\rangle\right] ,
\end{aligned}
\label{eq:noise_time_differentiation}
\end{equation}
where $\dot{\phi}_\alpha (x,t)\equiv \partial_t \phi_\alpha$ refers to the partial differentiation over time.
Following Eq.~\eqref{eq:noise_time_differentiation}, the noise can be obtained by evaluating the functions
\begin{equation}
\begin{aligned}
    & \mathcal{F}_{ud} (x_u,t_u;x_d, t_d) \equiv  \langle \left\{\phi_u (x_u, t_u), \phi_d (x_d,t_d) \right\}\big\rangle  -  2\big\langle \phi_u (x_u, t) \rangle \langle \phi_d (x_d, t_d) \big\rangle,\\
    & \mathcal{F}_{du} (x_d,t_d; x_u, t_u) \equiv  \langle \left\{\phi_d (x_d, t_d), \phi_u (x_u,t_u) \right\}\big\rangle  -  2\big\langle \phi_d (x_d, t_d) \rangle \langle \phi_u (x_u, t_u) \big\rangle,\\
    & S_{ud} (\omega;x_u,x_d) = \frac{\nu e^2}{8 \pi^2} \left[\int_{-\infty}^\infty dt e^{i\omega t} \mathcal{F}_{ud} (x_u,t;x_d, 0) + \int_{-\infty}^\infty dt e^{-i\omega t} \mathcal{F}_{du} (x_d,0; x_u, t) \right].
\end{aligned}
    \label{eq:f_function}
\end{equation}
In this section, we follow Ref.~\cite{DolciniPRB05}, and evaluate Eq.~\eqref{eq:f_function} with the generating functional method.
Within this method, two auxiliary fields, $\Lambda_u$ and $\Lambda_d$, are introduced as the conjugate variables of the field operators, $\phi_u$ and $\phi_d$, respectively.
With them, the average of a bosonic field operator, and the correlation function can be written as
\begin{equation}
\begin{aligned}
    & \big\langle  \phi_\alpha (x,t) \big\rangle = -\frac{i}{\sqrt{2}} \left( \frac{\delta Z}{\delta \Lambda_\alpha (x,t)} \right)\Bigg|_{\Lambda_u = \Lambda_d =0},\\
    & \big\langle \left\{ \phi_\alpha (x,t) \phi_{\alpha'} (x',t') \right\} \big\rangle - 2 \big\langle  \phi_\alpha (x,t) \big\rangle \big\langle \phi_{\alpha'} (x',t') \big\rangle
    =  \left( \frac{\delta Z}{\delta \Lambda_{\alpha} (x,t)} \frac{\delta Z}{\delta \Lambda_{\alpha'} (x' , t')} - \frac{\delta^2 Z}{\delta \Lambda_{\alpha} (x,t) \delta \Lambda_{\alpha'} (x', t')} \right)\Big|_{\Lambda_u = \Lambda_d = 0},
\end{aligned}
\label{eq:functional_expectation}
\end{equation}
with the generating functional
\begin{equation}
\begin{aligned}
    Z[\Lambda_u, \Lambda_d] & =  \iint \mathcal{D} \phi_u^\pm \mathcal{D} \phi_d^\pm \exp \left\{- \mathcal{S}_{u}^0[\Lambda_u, \phi_u] \right\}\  \exp\left\{- \mathcal{S}_{d}^0[\Lambda_d, \phi_d] \right\}\  \exp\left\{- \mathcal{S}_\mathcal{T}^0[\phi_u, \phi_d] \right\}\\
    \mathcal{S}_{u}^0[\Lambda_u, \phi_u] & = \frac{1}{2} \iint d\mathbf{r}' d\mathbf{r}'' \sum_{\eta,\eta' =\pm} \phi_u^\eta (\mathbf{r}') \left( g_u^{-1} \right)^{\eta,\eta'} (\mathbf{r}' ; \mathbf{r}'') \phi_u^{\eta'} (\mathbf{r}'')\\
    & \times \exp\left\{ \sum_\eta i  \int d\mathbf{r} \left(  \frac{e\sqrt{\nu}}{2 \pi} \eta E_u (\mathbf{r}) \phi_u (\mathbf{r} ) + \frac{1}{\sqrt{2}} \Lambda_u(\mathbf{r}) \phi_u^\eta (\mathbf{r}) \right)  \right\},\\
    \mathcal{S}_{d}^0[\Lambda_d, \phi_d] & =  \frac{1}{2} \iint d\mathbf{r}' d\mathbf{r}'' \sum_{\eta,\eta' =\pm} \phi_d^\eta (\mathbf{r}') \left( g_d^{-1} \right)^{\eta,\eta'} (\mathbf{r}' ; \mathbf{r}'') \phi_d^{\eta'} (\mathbf{r}'') \\
    & \times \exp\left\{ \sum_\eta i  \int d\mathbf{r} \left(  \eta \frac{e\sqrt{\nu}}{2 \pi} E_d (\mathbf{r}) \phi_d (\mathbf{r} ) + \frac{1}{\sqrt{2}} \Lambda_d (\mathbf{r}) \phi_d^\eta (\mathbf{r}) \right)  \right\},\\
    \mathcal{S}_\mathcal{T}^0[\phi_u, \phi_d] & =i \sum_\eta \eta \int dt H_\mathcal{T} [\phi_u^\eta (0), \phi_d^\eta (0)] ,
\end{aligned}
\label{eq:generating_functional}
\end{equation}
where $E_\alpha (\mathbf{r}) = \partial_x V_\alpha (\mathbf{r})$ equals the electric drive in channel $\alpha$. ``Effective actions'' $\mathcal{S}_u^0$, $\mathcal{S}_d^0$ are the functions involving only the field and source operators of the edge $u$ and $d$, separately, and $\mathcal{S}_\mathcal{T}^0$ characterizes the tunneling between these two edges.
Here we address them as ``effective'', as they involve the source terms, $E_\alpha$ and $\Lambda_\alpha$, that couple linearly to the field operator. These linearly coupled terms are typically absent in free actions.
These effective actions involve auxiliary fields $\Lambda_u$, $\Lambda_d$, and the bosonic fields $\phi_u$, $\phi_d$. The latter are further functions of $\mathbf{r} = (x, t)$, referring to the position in space and time.
Here $g_u^{-1}$ and $g_d^{-1}$ refer to the inverse matrix of the non-interacting Green's functions $g_u$ and $g_d$ [cf. Eq.~\eqref{eq:gf_matrix}], of the edge $u$ and $d$, respectively, with $\eta$, $\eta'$ the Keldysh indexes.
As its another crucial feature, in Eq.~\eqref{eq:generating_functional} the temperature only appears in the free parts, $\mathcal{S}_u^0$ and $\mathcal{S}_d^0$, leaving the tunneling part, $\mathcal{S}_\mathcal{T}^0$ explicitly temperature-independent.
This feature is the direct consequence of the fact that we work with real time in Eq.~\eqref{eq:generating_functional}.
Indeed, when working instead with the imaginary time, the temperature inverse simply enters as the temperature inverse.
Of the latter case (which differs from our current consideration), the temperature dependence is then explicitly present in all three parts of the action, via the imaginary time.

To evaluate correlation functions of Eq.~\eqref{eq:functional_expectation}, we perform a gauge transformation for the bosonized field operators of channel $\alpha$~\cite{DolciniPRB05} (as the reminder, here the superscript $\pm$ labels out the Keldysh branch),
\begin{equation}
\begin{aligned}
    \phi_\alpha^+ & \to \phi_\alpha^+ + \left( g_\alpha^{++} - g_\alpha^{+-} \right) \frac{\sqrt{\nu} e}{2\pi } E_\alpha +  \left( g_\alpha^{++} + g_\alpha^{+-} \right) \frac{1}{\sqrt{2}} \Lambda_\alpha,\\
    \phi_\alpha^- & \to \phi_\alpha^- + \left( g_\alpha^{-+} - g_\alpha^{--} \right) \frac{\sqrt{\nu} e}{2\pi } E_\alpha +  \left( g_\alpha^{-+} + g_\alpha^{--} \right) \frac{1}{\sqrt{2}} \Lambda_\alpha,
\end{aligned}
\label{eq:gauge_transformation}
\end{equation}
to remove the dependencies of the free-parts (i.e., $\mathcal{S}_{u}^0$ and $\mathcal{S}_d^0$) of the action on the corresponding bosonic field.
As its advantage, by doing so, the partition function can be factorized into three parts, $Z = Z_u Z_d Z_\mathcal{T}$, with their explicit expressions,
\begin{equation}
\begin{aligned}
    Z_u [\Lambda_u,E_u] & = \exp \left\{ - \iint d\mathbf{r}' d\mathbf{r}'' \left[ \frac{1}{2} \Lambda_u (\mathbf{r}') g^\mathcal{K}_u (\mathbf{r}'; \mathbf{r}'') \Lambda_u (\mathbf{r}'') + \sqrt{\frac{\nu}{2}} \frac{e}{\pi} \Lambda_u (\mathbf{r}') g^\mathcal{K}_u (\mathbf{r}'; \mathbf{r}'') E_u(\mathbf{r}'') \right]  \right\},\\
    Z_d [\Lambda_d,E_d] & = \exp \left\{ - \iint d\mathbf{r}' d\mathbf{r}'' \left[ \frac{1}{2} \Lambda_d (\mathbf{r}') g^\mathcal{K}_d (\mathbf{r}'; \mathbf{r}'') \Lambda_d (\mathbf{r}'') + \sqrt{\frac{\nu}{2}} \frac{e}{\pi} \Lambda_d (\mathbf{r}') g^\mathcal{K}_d (\mathbf{r}'; \mathbf{r}'') E_d(\mathbf{r}'') \right]  \right\},\\
    Z_\mathcal{T} & =  \iint \mathcal{D} \phi_u^\pm \mathcal{D} \phi_d^\pm \Bigg\langle T_K \exp \left\{ -i \frac{e^*}{\pi l_c} \sum_\eta \eta \int dt H_\mathcal{T} ( \phi_u^\eta (0) + A_u^\eta [\Lambda_u], \phi_d^\eta (0) + A_d^\eta [\Lambda_d] ) \right\} \Bigg\rangle_0,
\end{aligned}
\label{eq:rotated_generating_functional}
\end{equation}
where $T_K$ refers to the Keldysh ordering, and the integral over bosonic fields is contained only by the tunneling part of the partition function, due to the independence of ``free-edge'' partition functions, $Z_u$ and $Z_d$, on bosonic fields $\phi_u$ and $\phi_d$. We stress, however, that strictly speaking, these ``free-edge'' partition functions do not really come from the free-edge action $\mathcal{S}_u^0$ and $\mathcal{S}_d^0$ of Eq.~\eqref{eq:generating_functional}.
Indeed, the separation is rather effectively achieved via the transformation defined in Eq.~\eqref{eq:gauge_transformation}.
Actually, after the transformation, the free-edge Green's functions in Eq.~\eqref{eq:rotated_generating_functional} return to those defined in Eq.~\eqref{eq:impurity_free_gfs}, instead of those within the matrix, Eq.~\eqref{eq:gf_rotations}.
Importantly, the above-mentioned decomposition is only meaningful when the temperatures and biases of both channels are well defined, i.e., not influenced by the equilibration and thermal diffusion at the tunneling QPC.
The tunneling Hamiltonian after the transformation is now a function of the electric gauge field
\begin{equation}
\begin{aligned}
    A_\alpha^\eta (\mathbf{r})
    & = i \frac{e\sqrt{\nu}}{2\pi } \int d\mathbf{r}' g^\mathcal{R}_\alpha (\mathbf{r}; \mathbf{r}') E_\alpha (\mathbf{r}') + i \frac{1}{\sqrt{2}} \int d\mathbf{r}' \left[ g^\mathcal{K}_\alpha (\mathbf{r}; \mathbf{r}') + \eta g^\mathcal{A}_\alpha (\mathbf{r}; \mathbf{r}') \right] \Lambda_\alpha (\mathbf{r}'),
\end{aligned}
\label{eq:rotated_generating_functional}
\end{equation}
which is related to both the applied electric field $E_\alpha$ and the auxiliary field $\Lambda_\alpha$.
Physically, with the transformation, Eq.~\eqref{eq:gauge_transformation}, the voltage drive, initially given by Eq.~\eqref{eq:hbias}, becomes effectively absorbed by the coupling amplitude $\xi$, i.e., $\xi \to \xi  \exp \left\{ i e^* \int_{-\infty}^t dt' [V_u (t') - V_d(t')] \right\}$.
This is a famous trick taken from earlier references, e.g., \cite{DolciniPRB05,BenaSafiPBR07,ines_proceedings_2009,ines_philippe,Kane:1992,SaleurWeissPRB01}.

As the advantage of Eq.~\eqref{eq:rotated_generating_functional}, for positions out of the scattering area (i.e., being tunneling irrelevant), correlation functions between the bosonized field operators, i.e., $\phi_u$ and $\phi_d$, can be straightforwardly written as
\begin{equation}
\begin{aligned}
    \langle \phi_\alpha (\mathbf{r}_\alpha) \rangle & = -\frac{i}{\sqrt{2}} \left[ \frac{\partial Z_\alpha}{\partial \Lambda_\alpha(\mathbf{r}_\alpha)} + \frac{\partial Z_\mathcal{T}}{\partial \Lambda_\alpha(\mathbf{r}_\alpha)} \right]\Bigg|_{\Lambda_u = \Lambda_d =0}\\
    & = \frac{e\sqrt{\nu}}{2\pi} \int d\mathbf{r} g^\mathcal{R}_\alpha (\mathbf{r}_\alpha ; \mathbf{r}) E_\alpha (\mathbf{r}) + i\frac{1}{2 e} \sum_\eta \int dt \left( \eta g^\mathcal{K}_\alpha + g^\mathcal{A}_\alpha
 \right) (0, t; \mathbf{r}_\alpha) \langle \hat{I}_\mathcal{T}^\eta (0,t) \rangle,
\end{aligned}
\label{eq:phi_alpha_average}
\end{equation}
where, for later convenience, we define another space-time vector $\mathbf{r}_\alpha = (x_\alpha, t_\alpha)$.
Now we move to two-operator correlations.
Firstly, when performing the differentiation over the free part of the generating functional (i.e., that of two free edges), we obtain directly
\begin{equation}
\begin{aligned}
    & \frac{\partial (Z_u Z_d)}{\partial \Lambda_u(\mathbf{r}_u)} \frac{\partial (Z_u Z_d)}{\partial \Lambda_d(\mathbf{r}_d)} -\frac{\partial^2 (Z_u Z_d)}{\partial \Lambda_u(\mathbf{r}_u) \partial \Lambda_d(\mathbf{r}_d)}  \Bigg|_{\Lambda_u = \Lambda_d =0} = 0,\\
    & \frac{\partial Z_\alpha}{\partial \Lambda_\alpha(\mathbf{r})} \frac{\partial Z_\alpha}{\partial \Lambda_\alpha(\mathbf{r}')} -\frac{\partial^2 Z_\alpha}{\partial \Lambda_\alpha(\mathbf{r}) \partial \Lambda_\alpha(\mathbf{r}')}  \Bigg|_{\Lambda_u = \Lambda_d =0} = g^\mathcal{K}_\alpha (\mathbf{r}; \mathbf{r}'),
\end{aligned}
\label{eq:free_functional_differentiation}
\end{equation}
where the first and second lines are for the tunneling-free cross correlation and auto-correlations, respectively, i.e., that presented by Eq.~\eqref{eq:free_noises}.

After taking the tunneling at the QPC into consideration, relevant 
functional differentiation becomes complicated. Firstly, when considering the cross correlation, the differentiation over auxiliary fields becomes
\begin{equation}
\begin{aligned}
    &\frac{\partial^2 Z_\mathcal{T}}{\partial \Lambda_u (\mathbf{r}_u) \partial \Lambda_d (\mathbf{r}_d) } \Bigg|_{\Lambda_u=\Lambda_d=0} \!\!\!\! = \frac{i}{2}  \int dt \sum_\eta \left( \eta g^\mathcal{K}_u + g^\mathcal{R}_u  \right) (\mathbf{r}_u; 0, t) \left( \eta g^\mathcal{K}_d + g^\mathcal{R}_d  \right) (\mathbf{r}_d; 0, t)   \Bigg\langle \frac{\partial^2 H_\mathcal{T} \left[ \phi_u^\eta + A_{u}^\eta, \phi_d^\eta + A_{d}^\eta \right] }{\partial \phi_u (0, t) \partial \phi_d (0, t)} \Bigg\rangle\\
    & +\! \!\frac{1}{2}\!\! \iint \!dt' dt'' \sum_{\eta_1} \sum_{\eta_2} \! \left( \eta_1 g^\mathcal{K}_u \!+\! g^\mathcal{R}_u \right)\! (\mathbf{r}_u; 0, t') \left( \eta_2 g^\mathcal{K}_d \!\!+\!\! g^\mathcal{R}_d \right) \!(\mathbf{r}_d; 0, t'')  \Bigg\langle\! \frac{\partial H_\mathcal{T} \left[ \phi_u^{\eta_1} \!\!+\!\! A^{\eta_1}_{u}, \phi_d^{\eta_1}\! \!+\!\! A^{\eta_1}_{d} \right]}{\partial \phi_u (0, t')} \frac{\partial H_\mathcal{T} \left[ \phi_u^{\eta_2} \!\!+ \!\! A^{\eta_2}_{u}, \phi_d^{\eta_2} \!\!+\!\! A^{\eta_2}_{d} \right]}{\partial \phi_d (0, t'')}  \!\Bigg\rangle\\
    & = \frac{1}{2e^2} \iint dt' dt'' \sum_{\eta_1\eta_2} \eta_2 \left[ g^\mathcal{K}_u (\mathbf{r}_u; 0, t') g^\mathcal{R}_d (\mathbf{r}_d; 0, t') + g^\mathcal{R}_u (\mathbf{r}_u; 0, t') g^\mathcal{K}_d (\mathbf{r}_d; 0, t') \right] \langle \hat{I}^{\eta_1}_\mathcal{T} (t') \hat{I}^{\eta_2}_\mathcal{T} ( t'') \rangle \\
    & - \!\frac{1}{2e^2}\!\! \iint dt' dt'' \!\sum_{\eta_1\eta_2} \!\!\left\{ \eta_1 g^\mathcal{K}_u (\mathbf{r}_u; 0, t') g^\mathcal{R}_d (\mathbf{r}_d; 0, t'') \!\!+\!\! \eta_2 g^\mathcal{R}_u (\mathbf{r}_u; 0, t') g^\mathcal{K}_d (\mathbf{r}_d; 0, t'')  \!\!+\!\! g^\mathcal{R}_u (\mathbf{r}_u; 0, t') g^\mathcal{R}_d (\mathbf{r}_d; 0, t'') \right\} \langle \hat{I}^{\eta_1}_\mathcal{T} (t') \hat{I}^{\eta_2}_\mathcal{T} (t'') \rangle,\\
    & = \frac{1}{2e^2} \iint dt' dt'' \sum_{\eta_1\eta_2} \eta_2 \left[ g^\mathcal{K}_u (\mathbf{r}_u; 0, t') g^\mathcal{R}_d (\mathbf{r}_d; 0, t') + g^\mathcal{R}_u (\mathbf{r}_u; 0, t') g^\mathcal{K}_d (\mathbf{r}_d; 0, t') \right] \langle \hat{I}^{\eta_1}_\mathcal{T} (t') \hat{I}^{\eta_2}_\mathcal{T} ( t'') \rangle \\
    & -\! \frac{1}{2e^2}\!\! \iint dt' dt'' \!\sum_{\eta_1\eta_2}\!\! \left\{ \eta_1 g^\mathcal{K}_u (\mathbf{r}_u; 0, t') g^\mathcal{R}_d (\mathbf{r}_d; 0, t'') \!\!+\!\! \eta_2 g^\mathcal{R}_u (\mathbf{r}_u; 0, t') g^\mathcal{K}_d (\mathbf{r}_d; 0, t'')  \!\!+\!\! g^\mathcal{R}_u (\mathbf{r}_u; 0, t') g^\mathcal{R}_d (\mathbf{r}_d; 0, t'') \right\} \langle \hat{I}^{\eta_1}_\mathcal{T} (t') \hat{I}^{\eta_2}_\mathcal{T} (t'') \rangle \\
    & = \frac{1}{e^2} \iint dt' dt'' \left[ g^\mathcal{K}_u (\mathbf{r}_u; 0, t') g^\mathcal{R}_d (\mathbf{r}_d; 0, t') + g^\mathcal{R}_u (\mathbf{r}_u; 0, t') g^\mathcal{K}_d (\mathbf{r}_d; 0, t') \right] \theta (t'-t'') \Big\langle \left[ \hat{I}_\mathcal{T} (t'), \hat{I}_\mathcal{T} ( t'') \right] \Big\rangle \\
    & - \frac{1}{e^2} \iint dt' dt'' \left\{ \left[- g^\mathcal{K}_u (\mathbf{r}_u; 0, t') g^\mathcal{R}_d (\mathbf{r}_d; 0, t'') \theta (t'' - t')  + g^\mathcal{R}_u (\mathbf{r}_u; 0, t') g^\mathcal{K}_d (\mathbf{r}_d; 0, t'') \theta(t'-t'') \right] \Big\langle \left[ \hat{I}_\mathcal{T} (t'), \hat{I}_\mathcal{T} (t'') \right]\Big\rangle\right.\\
    &\left. + g^\mathcal{R}_u (\mathbf{r}_u; 0, t') g^\mathcal{R}_d (\mathbf{r}_d; 0, t'') \Big\langle \left\{\hat{I}_\mathcal{T} (t') \hat{I}_\mathcal{T} (t'') \right\}\Big\rangle \right\} ,
\end{aligned}
\label{eq:cross_functional_differentiation}
\end{equation}
where $\theta (t)$ is a step function (in time), and in the second equality, we have removed terms that are canceled out after summation over $\eta_1$ and $\eta_2$.
Starting from the third line, we are using Eq.~\eqref{eq:iud_general}, where the tunneling current operator is defined unambiguously in terms of the differentiation of $H_\mathcal{T}$, over the bosonic operators.
In addition, in comparison to that of Ref.~\cite{DolciniPRB05}, an extra overall minus sign is generated, due to differentiation over different phase factors, $\phi_u$ and $\phi_d$.
During the derivation, we have also used the identity
\begin{equation}
    \Bigg\langle \frac{\partial^2 H_\mathcal{T} \left[ \phi^\eta + A^\eta \right]}{\partial \phi_u (\mathbf{r}_u) \partial \phi_d (\mathbf{r}_d)} \Bigg\rangle = i \int dt' \sum_{\eta'} \eta' \Bigg\langle \frac{\partial H_\mathcal{T} [\phi_u^\eta + A_u^\eta,\phi_d^\eta + A_d^\eta]}{\partial \phi_u (\mathbf{r}_u)} \frac{\partial H_\mathcal{T} [\phi_u^{\eta'} + A_u^{\eta'} ,\phi_d^{\eta'} + A_d^{\eta'} ]}{\partial \phi_d (\mathbf{r}_d')} \Bigg\rangle,
\end{equation}
which can be obtained after functional integral by part.
To obtain the irreducible correlation, we further need
\begin{equation}
    \frac{\partial Z_\mathcal{T}}{\partial \Lambda_u (\mathbf{r}_u)} \frac{\partial Z_\mathcal{T}}{\partial \Lambda_d (\mathbf{r}_d)} \Bigg|_{\Lambda_u=\Lambda_d=0} = - \frac{2}{e^2 } \iint dt' dt'' g^\mathcal{R}_u (\mathbf{r}_u; 0, t') g^\mathcal{R}_d (\mathbf{r}_d; 0, t'') \langle \hat{I}_\mathcal{T} (t') \rangle \langle \hat{I}_\mathcal{T} (t'') \rangle ,
    \label{eq:cross_average_product}
\end{equation}
which produces the product of current averages.
Again, the extra overall minus sign of Eq.~\eqref{eq:cross_average_product} originates from the functional derivatives over both $\phi_u$ and $\phi_d$, which produce $I_\mathcal{T}$ and $-I_\mathcal{T}$, respectively.
In addition, here the argument of $\delta H_\mathcal{T}$ equals $\phi_u + A_u$, and thus contains the extra phase generated by the applied voltage bias.

Now, after the Fourier transform into the frequency domain, we can recover Eq.~\eqref{eq:cross_correlations} for the cross correlation function.
Actually, combining the last term of Eqs.~\eqref{eq:cross_functional_differentiation} and Eq.~\eqref{eq:cross_average_product} leads to $S^A_{ud}$ of Eq.~\eqref{eq:cross_correlations}. Terms from the last-but-one and last-but-two lines of Eq.~\eqref{eq:cross_functional_differentiation} instead lead to $S^C_{ud}$ of Eq.~\eqref{eq:cross_correlations}.

Likewise, we can derive Eq.~\eqref{eq:noise_relation} when $\alpha = \alpha'$, i.e., for auto correlations. In this case, Eqs.~\eqref{eq:cross_functional_differentiation} and \eqref{eq:cross_average_product} are correspondingly modified into
\begin{equation}
\begin{aligned}
    &\frac{\partial^2 Z_\mathcal{T}}{\partial \Lambda_\alpha (\mathbf{r}_\alpha) \partial \Lambda_\alpha (\mathbf{r}_\alpha) } \Bigg|_{\Lambda_u=\Lambda_d=0} \!\!\!\!\!\! 
    = -\frac{1}{e^2} \iint dt' dt'' \left[ g^\mathcal{K}_\alpha (\mathbf{r}_\alpha; 0, t') g^\mathcal{R}_\alpha (\mathbf{r}_\alpha; 0, t') \!\!+\!\! g^\mathcal{R}_\alpha (\mathbf{r}_u; 0, t') g^\mathcal{K}_\alpha (\mathbf{r}_\alpha; 0, t') \right] \theta (t'\!-\!t'') \Big\langle \left[ \hat{I}_\mathcal{T} (t'), \hat{I}_\mathcal{T} ( t'') \right] \Big\rangle \\
    & + \frac{1}{e^2} \iint dt' dt'' \left\{ \left[- g^\mathcal{K}_\alpha (\mathbf{r}_\alpha; 0, t') g^\mathcal{R}_\alpha (\mathbf{r}_\alpha; 0, t'') \theta (t'' - t')  + g^\mathcal{R}_\alpha (\mathbf{r}_\alpha; 0, t') g^\mathcal{K}_\alpha (\mathbf{r}_\alpha; 0, t'') \theta(t'-t'') \right] \Big\langle \left[ \hat{I}_\mathcal{T} (t'), \hat{I}_\mathcal{T} (t'') \right]\Big\rangle\right.\\
    &\left. + g^\mathcal{R}_\alpha (\mathbf{r}_\alpha; 0, t') g^\mathcal{R}_\alpha (\mathbf{r}_\alpha; 0, t'') \langle \left\{\hat{I}_\mathcal{T} (t'), \hat{I}_\mathcal{T} (t'') \right\}\rangle \right\} ,\\
    & \frac{\partial Z_\mathcal{T}}{\partial \Lambda_\alpha (\mathbf{r}_\alpha)} \frac{\partial Z_\mathcal{T}}{\partial \Lambda_\alpha (\mathbf{r}_\alpha)} \Bigg|_{\Lambda_u=\Lambda_d=0} = \frac{2}{e^2 } \iint dt' dt'' g^\mathcal{R}_\alpha (\mathbf{r}_\alpha; 0, t') g^\mathcal{R}_\alpha (\mathbf{r}_\alpha; 0, t'') \langle \hat{I}_\mathcal{T} (t') \rangle \langle \hat{I}_\mathcal{T} (t'') \rangle,
\end{aligned}
\label{eq:auto_functional_differentiation}
\end{equation}
which equals the opposite of Eqs.~\eqref{eq:cross_functional_differentiation} and \eqref{eq:cross_average_product}, after replacing the subscripts $u$ and $d$ by $\alpha$.
The difference, indicated by an overall minus sign induced by the differentiation over different phase factors, is perfectly captured by Eqs.~\eqref{eq:cross_correlations} and \eqref{eq:auto_noise_ac}.

\subsection{Effective tunneling within a finite-size area}
\label{app:finite_size_dyson}

In practical systems, tunneling at the QPC can be influenced by multiple factors, including e.g., a finite-size tunneling area, and influence from electron-electron and/or electron-phonon interactions.
Following Appendix~\ref{app:momentum_dependent}, all these practical factors can be 
(effectively) treated by the tunneling Hamiltonian Eq.~\eqref{eq:effective_tunneling_hamiltonian}, which has a position-dependent tunneling amplitude $\xi^\text{eff} (x,x')$, with $x$ and $x'$ labeling the positions (where tunneling occurs) of channel $u$ and $d$, respectively.
As the reminder, here we chose the convention within which $x$ and $x'$ increase towards the downstream direction, in both channels.
For later convenience, here we (without the loss of generality) further assume that $\xi^\text{eff} (x,x')$ is only finite when $ x_0^L < x, x' < x_0^R$ (with $x^R_0 $ at the downstream, as the reminder, following the definition), and equals zero otherwise.
In reality, of course $\xi^\text{eff} (x,x')$ varies continuous within the tunneling area. We can nevertheless choose $|x_0^L|$ and $|x_0^R|$ to be large enough, such that tunneling at points beyond is negligible.
In addition, anyons outside this area are assumed to be free from interactions [cf. Fig.~\ref{fig:momentum_dependent_transmission}(a)].
Before moving to derivation details, we remark that for systems with a finite-size tunneling area, Klein factors should be included, following e.g., Refs.~\cite{Safi:2001,Guyon:2002,Kane:2003,LawFeldmanGefenPRB06}.
In the derivations that follow, 
Klein factors are thus included in Eq.~\eqref{eq:current_density_operators}, when defining the density of tunneling current operators.
Nevertheless, as will be shown shortly, these Klein factors will not modify the final results.

To begin with, as in the ideal case, we decompose the action into three parts: two describing the free upper and lower edges, and one describing the transmission region. Since the free-edge parts do not involve inter-edge tunneling, even in more realistic situations, their corresponding actions, $Z_u$ and $Z_d$, remain identical to Eq.~\eqref{eq:rotated_generating_functional} for the ideal case.
This decomposition, however, is physically meaningful only if the temperature and voltage bias are well defined throughout the transmission region. Consequently, the derivations presented in the remainder of this section apply only when the thermal equilibration and diffusion have a negligible effect on $Z_u$ and $Z_d$.

When tunneling Hamiltonian contains a position-dependent effective tunneling amplitude, the functional differentiation of Eqs.~\eqref{eq:cross_functional_differentiation} and \eqref{eq:cross_average_product}, corresponding to cross correlation, becomes modified into
\begin{equation}
\begin{aligned}
    &\frac{\partial^2 Z_\mathcal{T}}{\partial \Lambda_u (\mathbf{r}_u) \partial \Lambda_d (\mathbf{r}_d) } \Bigg|_{\Lambda_u=\Lambda_d=0} \!\!\!\! = \frac{1}{e^2} \iint dt' dt'' \iint dx' dx''\\
    &\left( \left[ g^\mathcal{K}_u (\mathbf{r}_u; x', t') g^\mathcal{R}_d (\mathbf{r}_d; x'', t') + g^\mathcal{R}_u (\mathbf{r}_u; x', t') g^\mathcal{K}_d (\mathbf{r}_d; x'', t') \right] \theta (t'-t'') \Big\langle \left[ \hat{i}^\text{eff}_{u,\mathcal{T}} (t',x'), \hat{i}^\text{eff}_{d,\mathcal{T}} ( t'',x'') \right] \Big\rangle \right. \\
    & - \left\{ \left[- g^\mathcal{K}_u (\mathbf{r}_u; x', t') g^\mathcal{R}_d (\mathbf{r}_d; x'', t'') \theta (t'' - t') \! + \! g^\mathcal{R}_u (\mathbf{r}_u; x', t') g^\mathcal{K}_d (\mathbf{r}_d; x'', t'') \theta(t'-t'') \right] \Big\langle \left[ \hat{i}^\text{eff}_{u,\mathcal{T}} (t',x'), \hat{i}^\text{eff}_{d,\mathcal{T}} (t'',x'') \right]\Big\rangle\right.\\
    &\left. \left. + g^\mathcal{R}_u (\mathbf{r}_u; x', t') g^\mathcal{R}_d (\mathbf{r}_d; x'', t'') \Big\langle \left\{\hat{i}^\text{eff}_{u,\mathcal{T}} (t',x') ,\hat{i}^\text{eff}_{d,\mathcal{T}} (t'',x'') \right\}\Big\rangle \right\} \right),\\
    & \frac{\partial Z_\mathcal{T}}{\partial \Lambda_u (\mathbf{r}_u)} \frac{\partial Z_\mathcal{T}}{\partial \Lambda_d (\mathbf{r}_d)} \Bigg|_{\Lambda_u=\Lambda_d=0} = -\frac{2}{e^2 } \iint dt' dt'' \iint dx' dx'' g^\mathcal{R}_u (\mathbf{r}_u; x', t') g^\mathcal{R}_d (\mathbf{r}_d; x'', t'') \Big\langle \hat{i}^\text{eff}_{u,\mathcal{T}} (t',x') \Big\rangle \Big\langle \hat{i}_{d,\mathcal{T}}^\text{eff} (t'',x'') \Big\rangle ,
\end{aligned}
\label{eq:cross_functional_modified}
\end{equation}
where, different from the ideal situation (where $x_u, x_d$ require only to be positive), here both $x_u$ and $x_d$ are assumed to be large enough, to stay out of the interacting area.
Otherwise the theoretically obtained results can be significantly different from that obtained from realistic experimental measurements.
In Eq.~\eqref{eq:cross_functional_modified} we have further defined the tunneling current ``density'' operators (notice $x_0^R > x'$, for positions $x'$ inside the tunneling area)
\begin{equation}
\begin{aligned}
    i^\text{eff}_{u,\mathcal{T}} (x') & \equiv -\int dx \sqrt{\nu} \frac{\partial}{\partial \phi_u(x')} h_\mathcal{T} [\phi_u (x'), \phi_d (x)],\\
    i^\text{eff}_{d,\mathcal{T}} (x'') & \equiv -\int dx \sqrt{\nu} \frac{\partial}{\partial \phi_u(x)} h_\mathcal{T} [\phi_u (x), \phi_d (x'')],
\end{aligned}
\label{eq:current_density_operators}
\end{equation}
with integration over the positions of edges $d$ and $u$ introduced, in $i^\text{eff}_{u,\mathcal{T}}$ and $i^\text{eff}_{d,\mathcal{T}}$, respectively.
The definition of Eq.~\eqref{eq:current_density_operators} is introduced to capture the tunneling, $H_\mathcal{T} \equiv \int dx' dx'' h_\mathcal{T} [\phi_u (x'), \phi_d (x'')]$, that occur in a finite-size area, where $h_\mathcal{T} [\phi_u (x'), \phi_d (x'')]$ specifically refers to the tunneling between the positions $x'$ and $x''$,  of edges $u$ and $d$, respectively.

For the convenience of the readers, below we provide these operators in the weak-tunneling limit, i.e.,
\begin{equation}
\begin{aligned}
    i^\text{eff}_{u,\mathcal{T}} (x') & \equiv i e^* \int dx \xi^\text{eff}_\mathcal{T} (x',x) \: \left[ \psi^\dagger_d (x) \psi_u (x') - \psi^\dagger_u (x') \psi_d (x) \right],\\
    & = i e^* \int dx \frac{\xi^\text{eff}_\mathcal{T} (x',x) }{2 \pi l_c  }\:  \left\{ F_u (x') F_d^\dagger (x) e^{i\sqrt{\nu}\left[ \phi_u (x') - \phi_d (x ) \right]} + F_d (x) F_u^\dagger (x') e^{-i\sqrt{\nu}\left[ \phi_u (x') - \phi_d (x ) \right]} \right\} \\
    i^\text{eff}_{d,\mathcal{T}} (x'') & \equiv i e^* \int dx \xi^\text{eff}_\mathcal{T} (x,x'') \: \left[ \psi^\dagger_d (x'') \psi_u (x) - \psi^\dagger_u (x) \psi_d (x'') \right]\\
    & = i e^*\int dx'' \frac{\xi^\text{eff}_\mathcal{T} (x,x'') }{2 \pi l_c  }\: \left\{ F_u (x') F_d^\dagger (x'') e^{i\sqrt{\nu}\left[ \phi_u (x) - \phi_d (x'' ) \right]} + F_d (x'') F_u^\dagger (x) e^{-i\sqrt{\nu}\left[ \phi_u (x) - \phi_d (x'' ) \right]} \right\} ,
\end{aligned}
\label{eq:current_density_operators}
\end{equation}
corresponding to the \textit{total} current density received at the position $x'$ in channel $u$, and left from the position $x''$ in channel $d$, respectively, within the tunneling region.
Notice that operators of Eq.~\eqref{eq:current_density_operators} can also describe the tunneling towards the opposite direction, after introducing an extra minus sign.
Notice that in Eq.~\eqref{eq:current_density_operators}, an extra position-dependent phase factor should be included for tunneling towards (or from) different positions.
For simplicity, here we choose to include this phase factor into the definition of $\xi^\text{eff}_\mathcal{T}$.
Crucially, here we stress that
Eq.~\eqref{eq:cross_functional_modified} is non-perturbative, though the special expression, Eq.~\eqref{eq:current_density_operators}, strictly speaking requires the dependence of higher-order tunneling amplitudes on the leading-order one.

\begin{figure}
    \centering
    \includegraphics[width=0.95\linewidth]{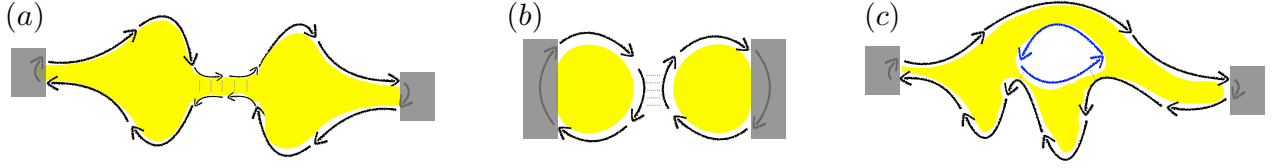}
    \caption{Topologically distinct FQH bars, with the bulk of the Hall bar highlighted in yellow.
    Their edge states are injected and received via the source and drain, respectively, that are both represented by the gray boxes.
    (a) The setup under current consideration, where the finite-size tunneling area (indicated by the gray dashed lines) shares the filling fraction of the bulk, and thus allows quasiparticles to transport through.
    (b) The setup where the QPC is topologically trivial (in contrast to the Hall bar indicated by the yellow area), and thus allows only electrons to tunnel through.
    Despite of the difference of the involved tunneling particles, both (a) and (b) contain only edges that travel clockwise.
    It is thus unambiguous to define upstream and downstream directions following Ref.~\cite{Safi:2001}, after including the states within two contacts.
    Klein factors are thus not necessary in these two setups.
    (c) Another setup (considered by e.g., Ref.~\cite{LawFeldmanGefenPRB06}) that contains a vacuum space (the white area encircled by the blue arrows) within the Hall bar.
    The inner edge, highlighted by the blue arrows, transport counter-clockwise, in contrast to that of the outer edge.
    Of this case, Ref.~\cite{Safi:2001} fails to apply, as the blue and black edges, transport counter-clockwise and clockwise, respectively, can not be combined and redefined into a single edge state.
    Of this case, Klein factors must be assigned to different tunneling QPCs~\cite{LawFeldmanGefenPRB06}.
    }
    \label{fig:qpc_klein_factor}
\end{figure}

In Eq.~\eqref{eq:current_density_operators}, $F_u$ and $F_d$ are two Klein factors that are assigned to the channel $u$ and $d$, respectively.
Following Refs.~\cite{Safi:2001,Guyon:2002}, they can be chosen to obey the anyonic braiding relation,
\begin{equation}
\begin{aligned}
    F_u F_d = e^{i\pi\nu} F_d F_u, \quad F_u^\dagger F_d = e^{-i\pi\nu} F_d F_u^\dagger.
    \label{eq:klein_factors}
\end{aligned}
\end{equation}
Within the main body of this work [corresponding to the setup of Fig.~\ref{fig:qpc_klein_factor}(a)], we choose to avoid these Klein factors [cf. Eq.~\eqref{eq:bosonization}] following Ref.~\cite{Safi:2001}, by combining two edges, $u$ and $d$ of Fig.~\ref{fig:model}, into a single one, after defining either $u$ or $d$ as at the downstream direction of the other one.
This treatment is actually valid mathematically, following discussions introduced in Ref.~\cite{Safi:2001}.
It, importantly, remains applicable to the setup of Fig.~\ref{fig:qpc_klein_factor}(b). Indeed, within this setup, all edges, like that of Fig.~\ref{fig:qpc_klein_factor}(a), transport clockwise, such that defining the relative upstream direction between different edges becomes unambiguous.
Treatment above, being applicable to the setups of Fig.~\ref{fig:qpc_klein_factor}(a) and (b), however fails to apply to that of Ref.~\cite{LawFeldmanGefenPRB06} (known as the Mach-Zehnder setup), where the bulk of the Hall bar contains an empty space [cf. the white area in the middle of Fig.~\ref{fig:qpc_klein_factor}(c)].
Here, the inner edge state (indicated by the blue arrows) transports counter-clockwise, in contrast to that of the outer edge state (the black arrows), such that it becomes impossible to define either the inner or the outer edge as the relatively upstream one.
Indeed, as shown by Ref.~\cite{LawFeldmanGefenPRB06}, Klein factors of this case have a significant influence on the observables.
Nevertheless, we stress that when evaluating the explicit expressions of tunneling current and noises, Klein factors, which will possibly influence the final outcome, should be involved (as that of Ref.~\cite{LawFeldmanGefenPRB06}).
Our result however indicates the irrelevance of these Klein factors to the non-perturbative fluctuation relations [cf. Eq.~\eqref{eq:noise_conductance_relation_brief}], where explicit expressions of noises and charge conductance are not necessarily requested: featuring the versatility of our fluctuation relations.

Following similar treatments of Eq.~\eqref{eq:cross_functional_modified}, Eq.~\eqref{eq:auto_functional_differentiation}, related to the evaluation of auto correlations, becomes modified into
\begin{equation}
\begin{aligned}
    &\frac{\partial^2 Z_\mathcal{T}}{\partial \Lambda_\alpha (\mathbf{r}_\alpha) \partial \Lambda_\alpha (\mathbf{r}_\alpha) } \Bigg|_{\Lambda_u=\Lambda_d=0} \!\!\!\! = -\frac{1}{e^2} \iint dt' dt'' \iint dx' dx''\\
    &\left( \left[ g^\mathcal{K}_\alpha (\mathbf{r}_\alpha; x', t') g^\mathcal{R}_\alpha (\mathbf{r}_\alpha; x'', t') + g^\mathcal{R}_\alpha (\mathbf{r}_\alpha; x', t') g^\mathcal{K}_\alpha (\mathbf{r}_\alpha; x'', t') \right] \theta (t'-t'') \Big\langle \left[ \hat{i}^\text{eff}_{\alpha,\mathcal{T}} (t',x'), \hat{i}^\text{eff}_{\alpha,\mathcal{T}} ( t'',x'') \right] \Big\rangle \right. \\
    & - \left\{ \left[- g^\mathcal{K}_\alpha (\mathbf{r}_\alpha; x', t') g^\mathcal{R}_\alpha (\mathbf{r}_\alpha; x'', t'') \theta (t'' - t') \! + \! g^\mathcal{R}_\alpha (\mathbf{r}_\alpha; x', t') g^\mathcal{K}_\alpha (\mathbf{r}_\alpha; x'', t'') \theta(t'-t'') \right] \Big\langle \left[ \hat{i}^\text{eff}_{\alpha,\mathcal{T}} (t',x'), \hat{i}^\text{eff}_{\alpha,\mathcal{T}} (t'', x'') \right] \Big\rangle\right.\\
    &\left. \left. + g^\mathcal{R}_\alpha (\mathbf{r}_\alpha; x', t') g^\mathcal{R}_\alpha (\mathbf{r}_\alpha; x'', t'') \Big\langle \left\{\hat{i}^\text{eff}_{\alpha,\mathcal{T}} (t',x') ,\hat{i}^\text{eff}_{\alpha,\mathcal{T}} (t'',x'') \right\}\Big\rangle \right\} \right),\\
    & \frac{\partial Z_\mathcal{T}}{\partial \Lambda_\alpha (\mathbf{r}_\alpha)} \frac{\partial Z_\mathcal{T}}{\partial \Lambda_\alpha (\mathbf{r}_\alpha)} \Bigg|_{\Lambda_u=\Lambda_d=0} = \frac{2}{e^2 } \iint dt' dt'' \iint dx' dx'' g^\mathcal{R}_\alpha (\mathbf{r}_\alpha; x', t') g^\mathcal{R}_\alpha (\mathbf{r}_\alpha; x'', t'') \Big\langle \hat{i}^\text{eff}_{\alpha,\mathcal{T}} (t',x') \Big\rangle \langle \hat{i}_{\alpha,\mathcal{T}}^\text{eff} (t'',x'') \Big\rangle .
\end{aligned}
\label{eq:auto_functional_modified}
\end{equation}
As the difference (from the ideal case), both Eqs.~\eqref{eq:cross_functional_modified} and \eqref{eq:auto_functional_modified} contain the integral over $x'$ and $x''$, i.e., positions in channels $u$ and $d$, respectively.

Following equations above, we can remarkably write down the generalized version of Eq.~\eqref{eq:noise_relation} [more specifically, Eqs.~\eqref{eq:cross_correlations} and \eqref{eq:auto_noise_ac}, for cross and auto correlations, respectively] as
\begin{equation}
\begin{aligned}
    \delta S_{\alpha\alpha'} (x,y,\omega) & = S^{A,\text{eff}}_{\alpha\alpha'} (x,y,\omega) + S^{C,\text{eff}}_{\alpha\alpha'} (x,y,\omega),\\
    S^{A,\text{eff}}_{\alpha\alpha'} (x,y,\omega) & = \left( 2\delta_{\alpha\alpha'} - 1 \right) \frac{\omega^2}{4\pi^2} \iint dx' dx'' g^\mathcal{R}_\alpha (x,x',\omega) g^\mathcal{R}_{\alpha'} (y,x'',-\omega) \int_{-\infty}^\infty dt e^{i\omega t} \Big\langle \left\{ \delta \hat{i}^\text{eff}_{\alpha,\mathcal{T}} (x', t), \delta \hat{i}^\text{eff}_{\alpha',\mathcal{T}} (x'', 0) \right\} \Big\rangle \\
    & = \left( 2\delta_{\alpha\alpha'} - 1 \right) \frac{\omega^2}{\pi} g^\mathcal{R}_\alpha (x,x_0^L,\omega) g^\mathcal{R}_{\alpha'} (y,x_0^L,-\omega) f^\text{eff}_{\alpha\alpha',A} ( \omega),\\
    S_{\alpha\alpha'}^{C,\text{eff}} (x,y,\omega) & = \left( 2\delta_{\alpha\alpha'} - 1 \right) \frac{\omega^2}{4\pi^2} \iint dx' dx'' \left[ g^\mathcal{K}_\alpha (x,x',\omega) g^\mathcal{R}_{\alpha'} (y,x'',-\omega) \int_0^\infty dt \left( e^{-i\omega t} - 1 \right)  \big\langle [\hat{i}^\text{eff}_{\alpha,\mathcal{T}} (x', t) , \hat{i}^\text{eff}_{\alpha',\mathcal{T}} (x'', 0)] \big\rangle \right. \\
    & \left.- g^\mathcal{R}_\alpha (x,x',\omega) g^\mathcal{K}_{\alpha'} (y,x'',-\omega) \int_0^\infty dt \left( e^{i\omega t} - 1 \right) \big\langle [\hat{i}^\text{eff}_{\alpha,\mathcal{T}} (x', t) , \hat{i}^\text{eff}_{\alpha',\mathcal{T}} (x'', 0)] \big\rangle \right]\\
    & = \left( 2\delta_{\alpha\alpha'} - 1 \right) \frac{\omega^2}{\pi}  \left[ g^\mathcal{K}_\alpha (x,x^L_0,\omega) g^\mathcal{R}_{\alpha'} (y,x^L_0,-\omega) f^\text{eff}_{\alpha\alpha',C} (-\omega)  - g^\mathcal{R}_\alpha (x,x',\omega) g^\mathcal{K}_{\alpha'} (y,x'',-\omega) f_{\alpha\alpha',C}^\text{eff} (\omega) \right],
\end{aligned}
\label{eq:noise_relation_extended}
\end{equation}
where the $\omega^2$ factor in each line cancels out the $\omega$ on the denominator of these Green's functions [cf. Eq.~\eqref{eq:free_greens_functions_space}], to avoid the zero-frequency singularity.
In Eq.~\eqref{eq:noise_relation_extended} $x$ and $y$ are positions at the downstream of the tunneling area, and the tunneling between two edges are included by two auxiliary functions,
\begin{equation}
\begin{aligned}
    f_{\alpha\alpha',A}^\text{eff} ( \omega) & = \frac{1}{4\pi}\int_{-\infty}^\infty dt e^{i\omega t} \iint dx' dx'' \Big\langle \left\{ \delta \hat{i}^\text{eff}_{\alpha,\mathcal{T}} (x', t), \delta \hat{i}^\text{eff}_{\alpha',\mathcal{T}} (x'', 0) \right\} \Big\rangle e^{i\nu \omega (x'-x'')/v}\\
    & \xrightarrow{\omega\to 0} \frac{1}{4\pi}\int_{-\infty}^\infty dt \Big\langle \left\{ \delta \hat{I}^\text{eff}_\mathcal{T} (t), \delta \hat{I}^\text{eff}_\mathcal{T} (0) \right\} \Big\rangle \\
    f_{\alpha\alpha',C}^\text{eff} ( \omega) & = \frac{1}{4\pi}\int_0^\infty dt \left( e^{i\omega t} - 1 \right) \iint dx' dx'' \big\langle [\hat{i}^\text{eff}_{\alpha,\mathcal{T}} (x', t) , \hat{i}^\text{eff}_{\alpha',\mathcal{T}} (x'', 0)] \big\rangle e^{i\nu \omega (x'-x'')/v}\\
    & \xrightarrow{\omega\to 0} \frac{\omega}{4\pi}\int_0^\infty dt \ t\  \big\langle [\hat{I}^\text{eff}_\mathcal{T} (t) , \hat{I}^\text{eff}_\mathcal{T} (0)] \big\rangle,
\end{aligned}
\label{eq:fa_fc_extended}
\end{equation}
where we take the zero-frequency limit, at the second line of each expression.
In Eqs.~\eqref{eq:noise_relation_extended} and \eqref{eq:fa_fc_extended}, the subscripts ``$A$'' and ``$C$'' highlight the involvement of the anti-commutator and commutator, respectively, of the tunneling current operator.
Here we keep the term proportional to $\omega$ of $f_{\alpha\alpha',C}^\text{eff} ( \omega)$, as the function $g^\mathcal{K}_\alpha (y,x'',-\omega)$, following Eq.~\eqref{eq:free_greens_functions_space}, diverges as $1/\omega$ (due to the $\coth$ function) in the zero-frequency limit.
In Eq.~\eqref{eq:fa_fc_extended}, the operator $\hat{I}^\text{eff}_\mathcal{T}$, defined as
\begin{equation}
\begin{aligned}
    \hat{I}^\text{eff}_\mathcal{T} (t) \equiv \int dx \hat{i}^\text{eff}_{\alpha,\mathcal{T}} (x,t) ,
\end{aligned}
\end{equation}
takes into the contribution to the tunneling current at all points, within the scattering area.
This operator is indeed the tunneling current operator through a finite-size QPC in the zero-frequency limit.
With its definition, the second line of $f_{\alpha\alpha',A}^\text{eff} ( \omega)$ in the $\omega \to 0$ limit is simply the tunneling current noise.

With the assistance of Eq.~\eqref{eq:fa_fc_extended} of two auxiliary functions, now Eq.~\eqref{eq:noise_relation_extended} for the correlation functions of the more practical situation is clearly equivalent to that of the ideal case, cf. Eqs.~\eqref{eq:cross_correlations} and \eqref{eq:auto_noise_ac}.
We have thus arrived at the conclusion that the non-equilibrium fluctuation relation, Eq.~\eqref{eq:noise_conductance_relation}, remains valid, even when the tunneling between two edges occurs within an area with finite size.
We stress again that, as discussed in Appendix~\ref{app:momentum_dependent}, this model with a finite-size tunneling area is actually effectively applicable to other systems.
This conclusion then further supports the validity of Eq.~\eqref{eq:noise_conductance_relation} in more complicated, and crucially, more realistic systems.
This conclusion, as discussed in the maintext, is not provided in previous researches, based on our knowledge.

\end{widetext}

\end{appendix}

\bibliography{biblio}

\end{document}